\documentclass[sigplan,10pt]{acmart}

\renewcommand\footnotetextcopyrightpermission[1]{} 
\acmConference[]{}{}{}

\setcopyright{none}

\usepackage{booktabs}   
\usepackage{subcaption} 

\newcommand{\XPTO}{ROLP}
\usepackage{amsmath}
\usepackage{algorithm}
\usepackage[noend]{algpseudocode}

\begin{document}

\title{\XPTO: Runtime Object Lifetime Profiling for Big Data Memory Management}


\author{Rodrigo Bruno}
\affiliation{
    \institution{INESC-ID / Instituto Superior T\'{e}cnico}            
  \city{University of Lisbon}
  \country{Portugal}                    
}
\email{rodrigo.bruno@tecnico.ulisboa.pt}          

\author{Duarte Patr\'{i}cio}
\affiliation{
  \institution{INESC-ID / Instituto Superior T\'{e}cnico}
  \city{University of Lisbon}
  \country{Portugal}
}
\email{duarte.carvalho@tecnico.ulisboa.pt}

\author{Jos\'{e} Sim\~{a}o}
\affiliation{
  \institution{INESC-ID / ISEL}
  \country{Portugal}
}
\email{jsimao@cc.isel.ipl.pt}

\author{Lu{i}s Veiga}
\affiliation{
  \institution{INESC-ID / Instituto Superior T\'{e}cnico}
  \city{University of Lisbon}
  \country{Portugal}
}
\email{luis.veiga@inesc-id.pt}

\author{Paulo Ferreira}
\affiliation{
  \institution{INESC-ID / Instituto Superior T\'{e}cnico}
  \city{University of Lisbon}
  \country{Portugal}
}
\email{paulo.ferreira@inesc-id.pt}


\begin{abstract}
Low latency services such as credit-card fraud detection and website targeted
advertisement rely on Big Data platforms (e.g., Lucene, Graphchi, Cassandra)
which run on top of memory managed runtimes, such as the JVM.
These platforms, however, suffer from unpredictable and unacceptably high
pause times due to inadequate memory management decisions (e.g., allocating
objects with very different lifetimes next to each other, resulting in memory
fragmentation). This leads to long and frequent application pause times,
breaking Service Level Agreements (SLAs). This problem has been previously
identified and results show that current memory management techniques are
ill-suited for applications that hold in memory massive amounts of middle to
long-lived objects (which is the case for a wide spectrum of Big Data
applications).

Previous works try to reduce such application pauses by allocating objects
off-heap or in special allocation regions/generations, thus alleviating
the pressure on memory management. However, all these solutions require
a combination of programmer effort and knowledge, source code access, or
off-line profiling, with clear negative impact on programmer productivity
and/or application performance.

This paper presents \XPTO, a runtime object lifetime profiling
system. \XPTO~profiles application code at runtime in order to identify which
allocation contexts create objects with middle to long lifetimes, given that
such objects need to be handled differently (regarding short-lived ones).
This profiling information greatly improves memory management decisions,
leading to long tail latencies reduction of up to 51\% for Lucene, 85\% for
GraphChi, and 60\% for Cassandra, with negligible throughput and memory
overhead. \XPTO~is implemented for the OpenJDK 8 HotSpot JVM and it does not
require any programmer effort or source code access.
\end{abstract}

\maketitle

\section{Introduction}
\label{sec:intro}

Big Data applications suffer from unpredictable and unacceptably high pause
times due to inadequate memory management decisions (e.g., allocating objects with very different lifetimes next to each other). This is the case of, for
example, credit-card fraud detection or website targeted advertisement systems
that rely on low-latency Big Data platforms (such as graph-based computing or
in-memory databases) to answer requests within a limited amount of time (usually
specified in Service Level Agreements, SLAs). Such pauses in these platforms delay application requests which can easily break SLAs.

The root cause of this problem has been previously identified
\cite{Rodrigo:2017:arxiv,Gidra:2012,Gidra:2013}
and can be decomposed in two sub-problems: i) Big Data platforms hold
large volumes of data in memory, and ii) data stays in memory for a long period
of time (from the Garbage Collector perspective), violating the widely accepted
assumption that most objects die young \cite{Ungar:1984,Jones:2008}.
Holding this middle to long-lived data in memory leads to an excessive Garbage Collection (GC) effort,
as described in Section \ref{sec:background}.
This results in long and frequent application pauses, which compromise application
SLAs, and demonstrates that current GC techniques, available in most industrial Java
Virtual Machines (such as OpenJDK HotSpot) are not suited for a wide spectrum of Big Data
Applications.

Recent works try to take advantage of application knowledge to better adapt Big
Data platforms to current GC techniques (in order to alleviate GC's work).
This can  be done either
through: i) manual refactoring of the application code, ii) adding code
annotations, or iii) static bytecode rewriting. The modified code reduces the
GC effort by either using off-heap memory,\footnote{Off-heap is a way to allocate objects
outside the scope of the GC. When using off-heap memory, the programmer is responsible
for allocating and collecting memory (by deallocating all previously allocated objects).}
or by redirecting allocations to scope limited
allocation regions \footnote{Scope limited allocation regions are used to allocate
objects which are automatically freed (with no GC effort) when the current scope
is terminated. Such objects cannot escape scopes.} or generations, leading
to reduced GC effort to collect memory. However, previous works have several drawbacks as they
require:
i) the programmer to change application code, and to know the internals of
GC to understand how it can be improved;
ii) source code access, which can be difficult if libraries or code inside the Java Development Kit needs to be modified; and
iii) workloads to be stable and known beforehand, since off-line profiling
performs code optimizations towards a specific workload.

Opposed to previous works, the goal of this work is to make GC pauses as short as
possible using online application profiling. The profiling technique must not have a negative impact
on the application throughput and on memory usage, and should not rely on any programmer effort or previous application
execution traces (such as those used for offline profiling).
Profiling information should be used to improve GC decisions, in particular, to allocate objects with
similar lifetimes close to each other.
Thus, \XPTO~strives to improve GC decisions by producing runtime object allocation statistics.
This profiling information must be accurate enough to avoid the need for any programmer
intervention or off-line application profiling. \XPTO~must reduce
application pause-times with negligible throughput or memory overhead, and it
must work as a pluggable component (which can be enabled or disabled at lunch time)
for current OpenJDK JVM implementations.

To attain such goals, \XPTO, a runtime object lifetime profiler (which runs inside the JVM),
maintains estimations of objects' lifetimes. This information is obtained by accounting the number of
both allocated and survivor objects. The number of allocated and survivor objects is kept in a global
table for each allocation context. This information can then be used by GC implementations to collocate
objects with similar lifetimes, thus reducing memory fragmentation and object promotion, the two
reasons for long tail latencies in current generational collectors.

Dynamically perceiving application allocation behavior at runtime is not trivial as no profiling information
from previous runs can be used, i.e., the profiler must only use information acquired at runtime.
This removes the need for
offline profiling and ensures that the profiling information is optimized for the current workload.


\XPTO~is implemented inside the OpenJDK 8 HotSpot JVM, one of the most widely used
industrial JVMs. In order to track allocation site allocations and allocation
contexts, profiling code is inserted during bytecode Just-In-Time compilation
\cite{Paleczny:2001}. To take advantage of profiling information, \XPTO~is integrated
with NG2C \cite{Rodrigo:2017:arxiv}, an N-Generational GC (based on Garbage First
\cite{Detlefs:2004}) that can allocate/pretenure objects into different
generation/allocation spaces.

\XPTO~supports any application that runs on top of the JVM
(i.e., it is not limited to the Java language) and users can benefit from
reduced application pauses with no developer effort or any need for off-line
profiling. As shown in the evaluation section (Section \ref{sec:eval}), when compared to other
approaches, \XPTO~greatly reduces application pause times (which results from
reduced object promotion and compaction). This is done with minimal throughput and
memory overhead.

In sum, the contributions of this work are the following:
\begin{itemize}
\item an online allocation profiling technique that can be used at runtime
to help generational garbage collectors taking better memory management
decisions, namely for object allocation;
\item the implementation of \XPTO, integrated with NG2C, for HotSpot, a widely
used production JVM. By implementing \XPTO~in such JVM, both researchers and companies
are able to easily test and take advantage of \XPTO;
\item an extensive set of performance experiments that compare \XPTO~with G1 (current
HotSpot default collector, described in detail in Section \ref{sec:background}), and NG2C
(that takes advantage of programmer knowledge combined with off-line profiling for allocating objects with similar lifetimes close to each other).
\end{itemize}


\section{Background}
\label{sec:background}
This section provides background on generational GC,
explaining its key insights, and why most implementations available in
industrial JVMs are not suited for low-latency Big Data platforms. NG2C, a pretenuring N-Generational collector, which \XPTO~builds upon is also discussed.

\subsection{Generational Garbage Collection}
Generational GC
is, nowadays, a widely used technique to
improve garbage collection \cite{Jones:2016}.
It is based on the observation that objects have
different lifetimes and, therefore, in order to optimize the collection process,
objects with shorter lifetimes should be collected more frequently than middle
to long-lived objects. To take advantage of this observation, the heap (memory
space where application objects live) is divided into generations (from youngest
to oldest). Most
generational collectors allocate all objects in the youngest generation and, as
objects survive collections, survivor objects are copied to older generations,
which are collected less frequently.

Most popular JVMs, and specifically the most recent HotSpot collector, called
Garbage First (G1) \cite{Detlefs:2004}, also takes advantage of the
weak generational hypothesis \cite{Ungar:1984}
which, as previously mentioned, states that most objects die young.
By relying on this hypothesis, G1 divides the heap into two generation:
young (where all objects are allocated), and old (where all objects that survive
at least one or more collections are copied to).

While this design works well for applications that follow the weak
generational hypothesis, it raises problems for applications that handle
many middle to long-lived objects. Such objects will be
promoted (i.e, moved from younger to older generations) and compacted through time (to
reduce memory fragmentation) until they become unreachable
(i.e., garbage). For example, many Big Data applications are not simple stream
processing platforms, but are more like in-memory databases where objects live
for a long time (from the GC perspective) and thus, the generational hypothesis
does not apply. In such cases, the cost of promoting and compacting
objects becomes prohibitive due to frequent and long application
pauses.

In other words, all generational collectors either assume or estimate the lifetime
of an object. Most collectors simply assume that all objects will
die young and pay the price of promoting and compacting survivor objects. As
we see next, some collectors, in particular NG2C, require application-level
knowledge to decide in which generation to allocate an object. By allocating
objects with similar lifetimes close to each other, directly into a specific generation
(and not assuming that all objects die young), NG2C is able to greatly reduce
memory fragmentation and therefore, the amount of promotion and compaction effort,
reducing the number and duration of application pauses.

\subsection{N-Generational Pretenuring}
\label{sec:background_ng2c}
NG2C \cite{Rodrigo:2017:arxiv} extends 2-generational collectors (which only contain two generations:
young and old) into an arbitrary number of generations. In addition,
it allows objects to be pretenured into any generation, i.e.,
objects can be allocated directly into any generation
(the idea is that objects with similar lifetimes are allocated in
the same generation).

To decide where to allocate objects, NG2C uses an external profiler, combined
with programmer effort and knowledge to annotate application code. Code annotations,
indicating which objects should go into which generation are then used by
the collector to decide where to allocate each object. As presented by the
authors, this leads to significant reduction in the amount of object promotion
and compaction, leading to reduced application pauses times.

However, in order to present good results, NG2C relies on off-line profiling to
extract the lifetime distribution for each allocation site and therefore,
estimate the lifetime of an object allocated in each allocation site. In
addition, the programmer must also use the output of the profiler to change
the application code.
As mentioned before, this introduces several problems:
i) the application needs to be profiled each time the application
is updated or the workload changes,
ii) it requires source code access (which is difficult if the code to change
resides in some external library or inside the JDK), and finally
iii) it requires programmer effort and knowledge to change the code correctly.

\section{Object Allocation Profiling}
\label{sec:asp}
This section describes \XPTO, an online object lifetime profiler that profiles application code in order to track application allocation patterns
(allowing the GC to allocate objects with similar lifetimes close to each other). \XPTO~has three main components:
i) a data structure that holds profiling data (i.e., object lifetimes per allocation context);
ii) code instrumentation during JIT (Just-In-Time) compilation to collect profiling information (as described next, \XPTO~instruments both allocation sites and method calls);
iii) internal JVM tasks that consolidate profiling information and suggest different generations to different allocation contexts whenever appropriate.
The next sections explain in detail these three components.

\begin{figure}[t]
\centering
\includegraphics[keepaspectratio,width=.2\textwidth]{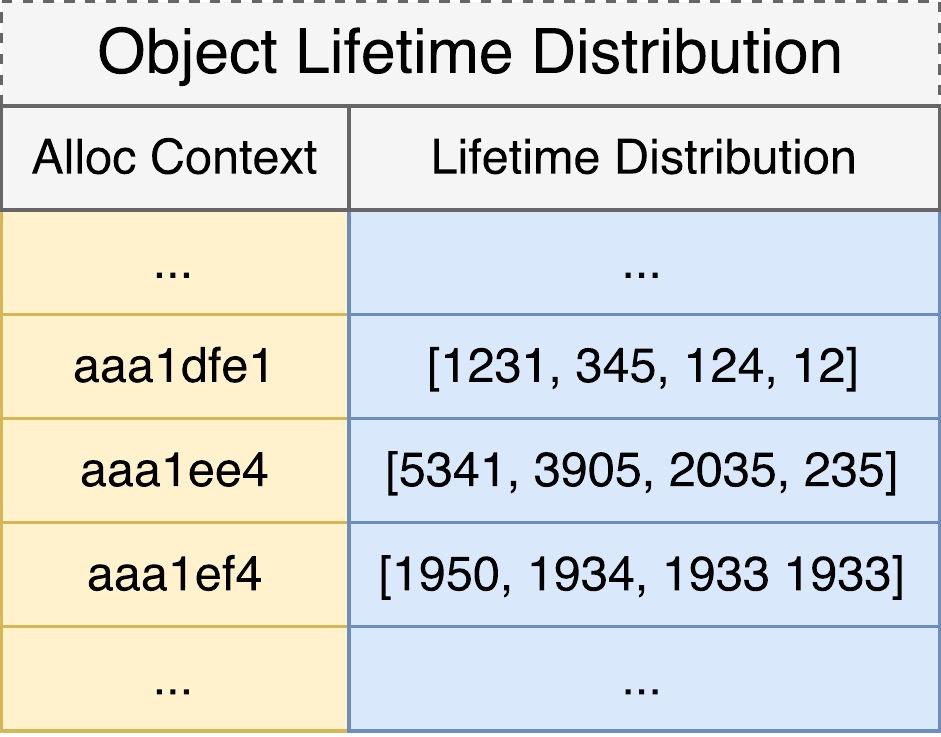}
\caption{Object Lifetime Distribution Table}
\label{fig:global-alloc-counter}
\end{figure}

\begin{figure}[t]
\centering
\includegraphics[keepaspectratio,width=.40\textwidth]{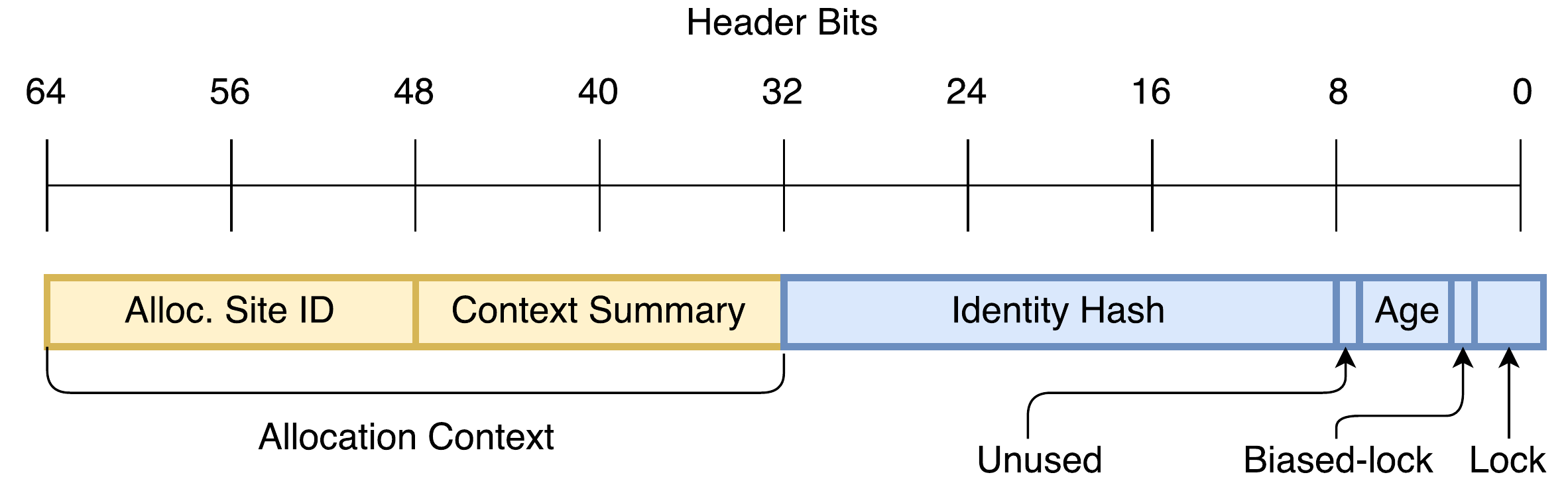}
\caption{\XPTO~Object Header in HotSpot JVM}
\label{fig:obj-header}
\end{figure}

\subsection{Allocation Context Lifetime Distribution Table}
\label{sec:asp_global_ds}
In \XPTO, each allocated application object is associated with an allocation context,
a 32 bit identifier (see Figure \ref{fig:obj-header}) composed by two 16 bits components (the size of each component
is configurable):
i) a context summary, and
ii) an allocation site identifier (obtained by hashing the method signature
concatenated with the line number of the allocation).
The context summary is a value that represents the current execution state while
the allocation site identifier represents the code location where the allocation
takes place. By using these two components, an allocation context not only represents
the code location where an allocation takes place but also the execution
state (method calls in particular) that led to the object allocation. In other words:
i) two objects allocated through different allocation sites will always have different
allocation contexts (because the allocation site identifier is different), and ii)
two objects allocated through the same allocation site will have different allocation
contexts if the context summary that led to the allocation is different.

To track the age of objects allocated through each allocation context, \XPTO~maintains
a global lifetime distribution table. This table contains, for each allocation context,
one lifetime distribution array (array of integers).
Each array position, from 0 to N (a configurable variable),
represents the number of objects that survived 0 to N collections. Figure
\ref{fig:global-alloc-counter} depicts an example of this table with N = 4.

The next sections describe how the context summary is managed and how the information in the object lifetime table is updated.

\subsection{Code Instrumentation during JIT Compilation}
\label{sec:asp_object_alloc}
Profiling code has a cost both in terms of time (to produce/insert and to execute)
and memory (increases the code size and requires more space to save profiling data).
In order to limit these overheads, ROLP instruments only highly executed/hot methods, thus
avoiding both time and memory overheads for methods that are not executed frequently.

In order to identify hot code locations, \XPTO~piggy-backs instrumentation
code on the Just-In-Time compilation process, which compiles frequently executed
methods (converting application bytecodes into native code).
In particular, \XPTO~intercepts the compilation/expansion of specific bytecodes
and inserts the needed profiling code.

The profiling code needs to perform three tasks:
i) update the thread-local context summary whenever the execution flow enters or leaves a method;
ii) increment the number of allocated objects in the corresponding allocation context (upon object allocation); and
iii) mark the allocated object with a corresponding allocation context (upon object allocation).
The next sections describe each of these tasks.

\subsubsection{Allocation Context Tracking}
The context summary is necessary to distinguish two
object allocations that, although using the same allocation site (i.e., the same
code location), use different call graphs to reach the allocation site.

To track allocation contexts, \XPTO~relies on two observations/requirements.
First, for allocation tracking purposes, it suffices that the context summary
differentiates (as much as possible) two different call graphs. However, the
details of the method calls that compose the call graph and their order
(i.e., which method call was executed before the other) is not required to be contained in the context summary.
Second, the context summary must be incrementally maintained as
the application execution goes through the call graph.

To fulfill these requirements, \XPTO~uses simple arithmetic operations (sum and
subtraction) to incrementally maintain a 32 bit thread local context summary. Thus,
before each method call, the thread local context summary is incremented with a unique
method call identifier/hash. The same value is subtracted when the execution exits the method.

As discussed next, the context summary is combined with the allocation site id to
generate the allocation context. Using both components is important to deal with
context summary collisions (which can happen if two different call paths
happen to generate the same context summary).
Hence, using both components,
the allocation context collision rate is reduced (as shown in Section \ref{sec:eval_profiling}).


\subsubsection{Updating the Number of Allocated Objects}
\begin{figure}[t]
\centering
\includegraphics[keepaspectratio,width=.45\textwidth]{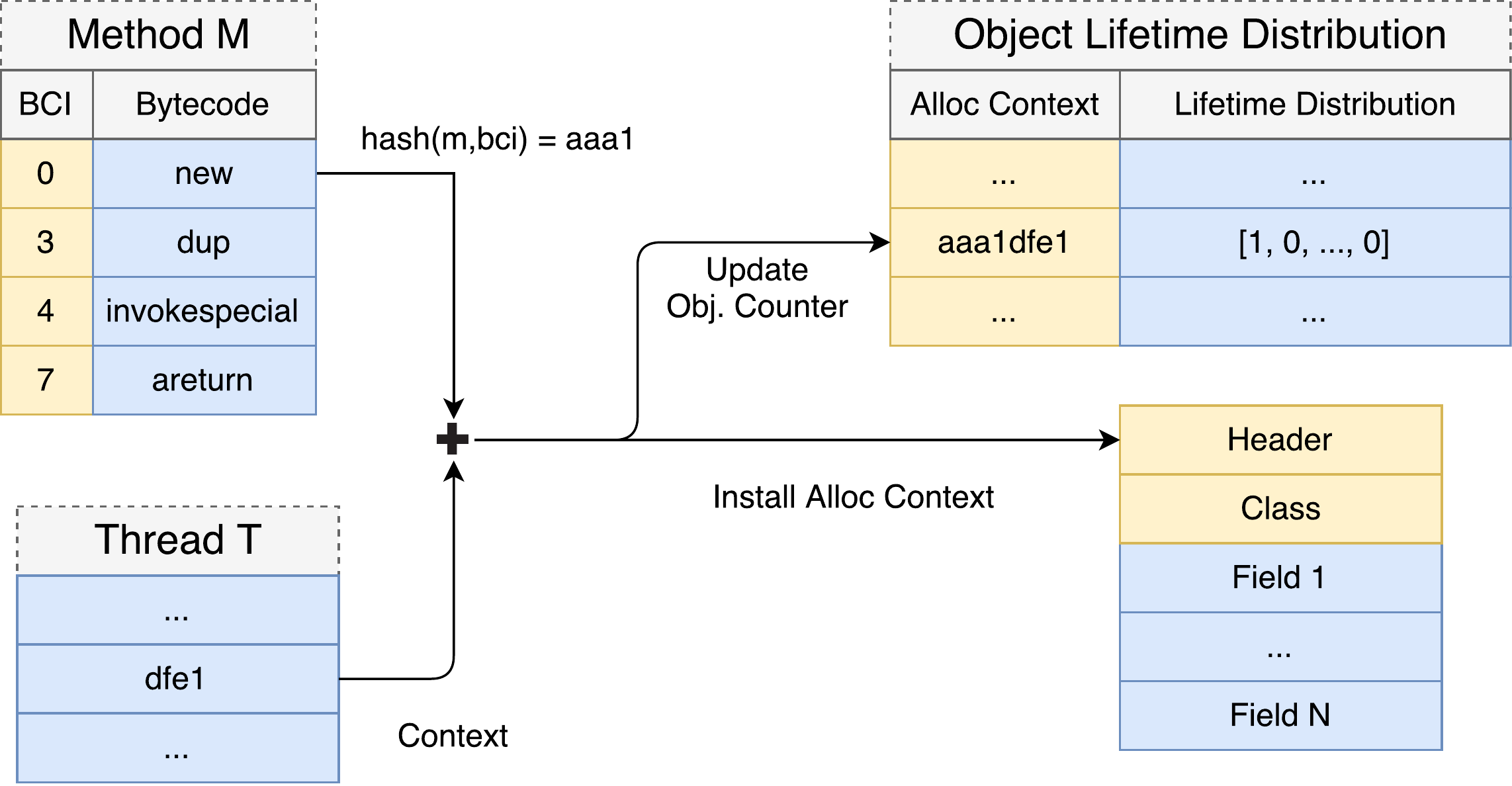}
\caption{Updating Object Counter and Header}
\label{fig:up_obj_alloc}
\end{figure}

The number of allocated objects per allocation context is maintained in the
object lifetime distribution table (see Figure \ref{fig:global-alloc-counter}).
This table contains
the numbers of objects that survived 0 to N collections. Thus, newly allocated
objects are accounted in the first position of the array associated to the
specific allocation context.

As depicted in Figure \ref{fig:up_obj_alloc}, upon each object allocation,
the allocation context is generated by concatenating
the allocation site identifier with the thread local context summary. Using
the allocation context, it is possible to access the corresponding entry in the
lifetime distribution table to increment the first position of the corresponding
array.

\subsubsection{Marking Objects with the Allocation Context}
Tracking object lifetimes cannot be done by only tracking the number of
allocations. It is also required to: i) tag objects allocated through each specific
allocation context, in order to ii) identify and trace back the correct allocation context
for each survivor object.

To identify the allocation context where an object was allocated, \XPTO~needs to
associate each object with one allocation context. Adding more
information to application objects (for example, increasing the header size)
is undesirable as it increases the memory footprint by adding extra bytes to
every object. Therefore, \XPTO~reuses spare bits that already exists in the
object header.

Figure \ref{fig:obj-header} presents the 64-bit object header used in the
HotSpot JVM. The first three bits (right to left) are used by the JVM for locking purposes,
followed by the age of the object (bits 3 to 6) which is also maintained by the
JVM. Bit number 7 is unused and bits 8 to 32 are used to store the object
identity hash, an object unique identifier.

As depicted in Figure \ref{fig:up_obj_alloc}, \XPTO~installs the allocation context
in the upper 32 bits of the 64-bit header used in HotSpot object
headers. These 32 bits are only used when the object is biased locked towards
a specific thread \footnote{Biased Locking is a locking technique available
for the HotSpot JVM which allows one to lock an object towards a specific thread.
It improves the object locking speed for the presumably most frequent scenario,
the object will only be locked by a single thread \cite{dice2010quickly}.} and using
them does not compromise the semantics of biased locks. Since \XPTO~installs the
allocation context upon object allocation, if the object becomes biased locked, the profiling
information will get overwritten. In addition, biased locking is controlled by the JVM using
a specific bit in the object header (bit number 3).

Using space dedicated for biased locks means that we might loose some profiling information.
However, through our experience, we argue that:
i) the number of biased locked objects in Big Data applications is not
significant;
ii) data objects are usually not used as locks (and therefore are not biased locked);
iii) not profiling non-data objects does not lead to a significant performance loss.

In sum, \XPTO~installs a 32 bit allocation context into each object
header. By doing this, \XPTO~is able to back trace the allocation
context for any object. The allocation context may become unusable if the corresponding
object becomes biased locked. In this situation, the object is not considerate for
updating the profiling information.

\subsection{Updating Object Lifetime Distribution}
\label{sec:asp_object_promo}
Objects' lifetime distribution is kept in the global table presented in
Figure \ref{fig:global-alloc-counter}. Objects that survived zero or more collections
(i.e., all allocated objects) are
accounted in the first position of the array.
Objects that survived one or more collections are accounted in the second
position of the array. This goes up to N collections, a number that can be
configurable at launch time.

Object lifetimes are measured in number of survived collections.
To do so, \XPTO~intercepts object collections and, for
each survivor object, it:
i) extracts the allocation context identifier (upper 32 bits of the corresponding
object's header),
ii) extracts the age (number of survived collections) of the corresponding object
(bits 3 to 6 in Figure \ref{fig:obj-header}),
iii) obtains the lifetime distribution array corresponding to the allocation
context, and
iv) increments the position indexed by the age of the corresponding object.

\begin{figure}[t]
\centering
\includegraphics[keepaspectratio,width=.3\textwidth]{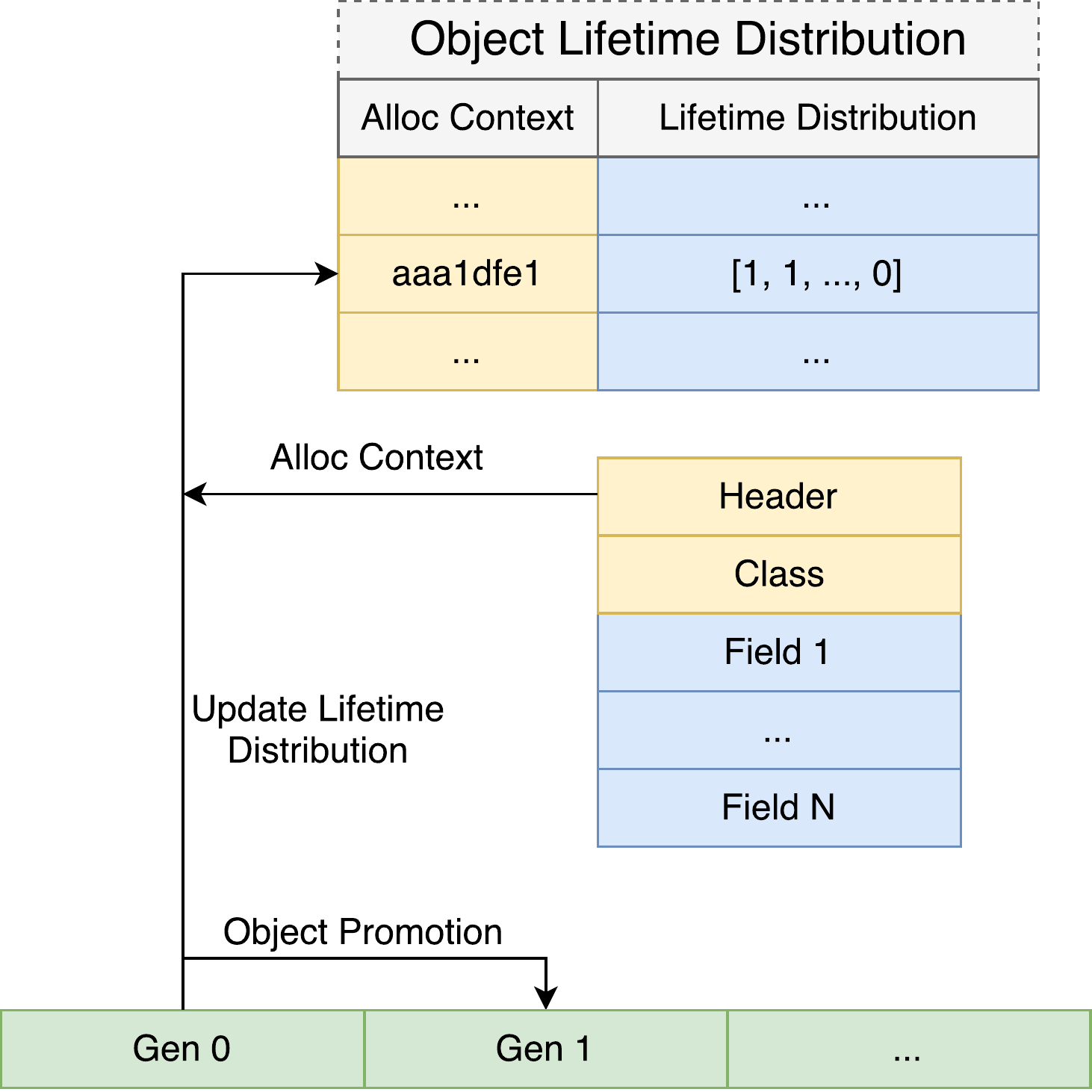}
\caption{Updating Object Promotion/Compaction}
\label{fig:up_obj_promo}
\end{figure}

This process is also depicted in Figure \ref{fig:up_obj_promo}.
By the end of each collection, the global table presented in Figure
\ref{fig:global-alloc-counter} contains, for every allocation context, the number
of objects that survived up to N collections. As discussed in the next section,
this information is used to improve GC decisions, namely the target generation
for object allocations.

\section{Dynamic N-Generational Pretenuring}
\label{sec:aop}
As described in previous sections, \XPTO~maintains lifetime information
for objects allocated through each allocation context. This information, as
described next, is used to determine in which generation each object should
be allocated, according to its allocation context.
In other words, this section describes how \XPTO~is integrated with NG2C
\cite{Rodrigo:2017:arxiv}, an N-Generational GC (based on G1) that allows objects
to be allocated in different generations. Note that, however, \XPTO~can be
integrated with any GC that supports pretenuring (i.e., allocating objects
directly in older generations/allocation spaces).

The result of the combination of \XPTO~with NG2C is a GC that is capable of
dynamically deciding where (i.e., in which generation/allocation space) to
allocate objects based on profiling information gathered at runtime. By doing
so, object promotion and compaction are greatly reduced, leading to reduced
number and duration of application pauses (as demonstrated in Section \ref{sec:eval}).
This can be achieved without off-line profiling and/or any programmer effort/knowledge.

The integration of \XPTO~with NG2C can be summarized in two points:
i) adding
an extra column in the object lifetime table (presented in Figure
\ref{fig:global-alloc-counter}) to contain the target generation for each
allocation context (and changing NG2C to use it instead of relying on code annotations), and
ii) determining how to properly update the
target generation taking into consideration the profiling information presented in the
object lifetime table.

To update the target generation taking as input the profiling information maintained in
the object lifetime table, \XPTO~uses an algorithm that updates the target generation, i.e.,
that decides in which generation objects are allocated (see Alg. \ref{alg:update_tgen}).
The algorithm starts by determining the current survivor
threshold, the number of collections that each object allocated in the young generation
must survive until it gets promoted/copied from the young generation into the old
generation. This limit is managed
by GC ergonomics (a set of
GC policies that control aspects related to heap and GC in general) and therefore,
changes over time.

For each object lifetime distribution array and corresponding target generation,
the algorithm described in Alg. \ref{alg:update_tgen} will determine if the
ratio between the number of promoted/copied and allocated objects is superior
to a predefined limit, \textit{INC\_GEN \_THRES}. If so, the target
generation for that specific allocation context is incremented. The threshold
defined by \textit{INC\_GEN\_THRES}, which has a default value that can be overridden
at JVM launch time.

As suggested by the algorithm, the target generation for an allocation context
is never decremented. Although this might seem counter intuitive, according to
our experience, having a large number of generations to hold many classes of
object lifetimes is better than spending time trying to determine if the target
generation should be decremented. In other words, there is no performance drawback
in letting allocation sites increment their target generation even if in the future,
the lifetime of objects allocated through a specific allocation context, sharing
the same generation, change. In this case, there will be some temporary fragmentation
(due to
objects with different lifetimes being in the same generation) but eventually objects
that live longer will push their allocation context into another generation.

Finally, to better support workload dynamics, the presented algorithm only
runs after a configurable number of collections (default value can be overridden
at JVM launch time, using the variable \textit{NG2C\_INC\_GEN\_FREQ}).
After each execution of the algorithm, all object lifetime table entries are reset to
zero. This provides a window during which the object lifetime table is updated
and allows the algorithm to take decisions based on fresh data (number of allocated
and promoted objects).

\begin{algorithm}[t]
\caption{Updating Target Generation}
\label{alg:update_tgen}
\begin{algorithmic}[1]
\Procedure{Update\_Target\_Generation}{}
\State $\textit{survivor\_threshold} \gets \text{GC ergonomics}$
\For{\textit{array}, \textit{target\_gen} \textbf{in} \textit{object lifetime table}}
    \State $\textit{allocated} \gets \textit{array}[0]$
    \If{\textit{N} > \textit{survivor\_threshold}}
        \State $\textit{promoted} \gets \textit{array}[\textit{survivor\_threshold}]$
    \Else
        \State $\textit{promoted} \gets \textit{array}[\textit{N}-1]$
    \EndIf
    \State $\textit{ratio} \gets \textit{promoted} / \textit{allocated}$
    \If{\textit{ratio} > \textit{INC\_GEN\_THRES}}
        \State \textit{target\_gen}++
    \EndIf
\EndFor
\EndProcedure
\end{algorithmic}
\end{algorithm}


\section{Implementation}

\label{sec:impl}
To implement \XPTO, the OpenJDK 8 HotSpot JVM was modified to automatically estimate
the object lifetime for objects allocated through each allocation context. To do so,
it needs to install profiling code and maintain several data structures. \XPTO~also
integrates with NG2C, which takes advantage of the profiling information
gathered by \XPTO~to automatically pretenure objects.

Since HotSpot is a highly optimized production JVM, new algorithms/techniques
must be implemented carefully so as not to break the JVM's performance.
This section describes some of \XPTO's implementation
details, in particular, the ones we believe to be important for realistically
implementing \XPTO~in a production JVM.

\subsection{Efficient Management of Profiling Structures}
\label{sec:efficient_management}
As illustrated in Figures~\ref{fig:up_obj_alloc} and~\ref{fig:up_obj_promo},
\XPTO~tracks allocation statistics regarding allocated
and promoted/compacted objects. This information must be updated by application
threads (to update the number of allocated objects) and GC worker threads
(to update the number of objects that survived collections). The next sections
describe some important implementation details for both of these steps.

\subsubsection{Updating Object Allocation}
In order to allow fast updates by application threads, two options were analyzed:
i) have a thread-local table, which periodically is used to update the global
table; ii) use a global table with no access synchronization (risking some increment misses).

\XPTO~uses the latter approach for three reasons: i) it is possible to write
more efficient native code (jitted code) because the address where the counter
(that needs to be incremented) resides is already known at JIT time; ii) it requires
less memory to hold profiling information; iii) the probability of loosing counter
increments is small. In other words, we
are trading performance for precision. However, according to our experience while
developing \XPTO, this loss of precision is not enough to change profiling decisions, i.e.,
the profiler takes the same decisions with and without synchronized counters.

\subsubsection{Updating Object Promotion/Compaction}
\label{sec:efficient_promotion}
GC worker threads must also update the global object lifetime distribution
table to account for objects that survive collections. However, opposed to
application threads, the contingency to access the global table is higher
since all worker threads may be updating the table at the same time during
a garbage collection. This would lead to significant loss of precision if no
synchronization takes place.
In order to avoid that, private tables (one for each GC worker thread)
containing only information regarding the objects promoted/compacted by a
particular worker thread are used. All these private tables are used to update the
global table
right after the current garbage collection finishes (more details in Section \ref{sec:limiting_gc_lat}).

\subsubsection{Object Lifetime Distribution Table Scalability}
\label{sec:efficient_table}
\XPTO~uses a global lifetime distribution table which is accessed every time an object is accessed very frequently.
In order to provide average constant time for insertion and search, this data structure is implemented as a hashtable

Another important concern is how large is the memory budget to hold this table
in memory. In the worst case scenario, and since the allocation context is a 32 bit
value, one could end up with a table with $2^{32}$ entries which would take, assuming
a life distribution array with 8 slots (i.e. N = 8), each of which using 4 bytes, 32 * $2^{32}$ bytes (approximately 128GB).

In order to reduce the memory footprint of this table, \XPTO~starts by grouping all
information regarding each allocation site (and disregarding the context summary).
Thus, in the beginning, the table will only grow up to $2^{16}$ entries (since there
are only $2^{16}$ possible allocation site ids using a 16 bit allocation site identifier), resulting is approximately 2MB
of memory.

In sum, by default, the object lifetime table groups all allocations and object
promotions/compactions from a particular allocation site in the same entry (regardless
of the context summary). To decide if a specific allocation site should start tracking
allocations for each context summary (and thus split into multiple entries, one for each
allocation context), \XPTO~further extends the algorithm presented in Alg.
\ref{alg:update_tgen}. The extended version not only increases the target
generation if the ratio between survivor and allocated objects is above
\textit{INC\_GEN\_THRES}, but also decides to
start tracking contexts for the particular allocation site if the ratio is above
EXPAND\_CTX (a configurable variable defined at launch time).
\begin{equation}
  0 < \textit{EXPAND\_CTX} < \textit{INC\_GEN\_THRES} < 1
\end{equation}
In practice, both variables (\textit{EXPAND\_CTX} and \textit{INC\_GEN \_THRES})
should be configured
such that Eq. 1 is verified. Using such configuration, allocation sites that allocate
objects with short lifetimes will never neither expand their contexts nor increment their target generation. If the ratio of survivor objects is above \textit{INC\_GEN\_THRES},
the target generation will be incremented for all objects allocated for that particular
allocation site. If the ratio falls between both variables, it means that objects
allocated through the same allocation site have different lifetimes. Only in this
case, contexts will be expanded. This results in a small extra memory cost to track
allocations per allocation context.

\subsection{Reducing Profiling Effort with Static Analysis}
\label{sec:static_analysis}
In order to successfully track allocation contexts, \XPTO~profiles
method calls. Before and after each method call, the context
summary of the executing thread is updated. This is a costly technique
and therefore, only method calls that are important to characterize the allocation
context should be profiled.

In order to reduce the number of profiled method calls, \XPTO~takes advantage of static
bytecode analysis (that can be performed offline, for example, after each release build)
to determine:
i) method calls that will never lead to object allocations (for example, most
getter and setter methods);
ii) methods calls that are more than MAX\_ALLOC\_FRAME (a variable defined at launch
time) frames of distance from any object allocation.

Combining these techniques with the fact that only JIT compiled method calls
are profiled, the number of profiled method calls is significantly reduced,
thus allowing a low throughput overhead implementation for context tracking.

By default, \XPTO~runs the static analyzer for the whole application (provided
through the classpath JVM argument). However, for very large applications,
there is still (despite the optimizations previously described) a potentially
large number of methods to be profiled. In order to reduce the number of profiled
methods, and therefore the profiling overhead, developers can explicitly select
which packages should be statically analyzed. According to our experience,
limiting the statical analysis to packages containing the application's data
structures leads to significant overhead reduction.

\subsection{Pushing Target Generation Updates out of Safepoints}
\label{sec:limiting_gc_lat}
Periodically, the algorithm described in Alg. \ref{alg:update_tgen}
must be executed in order to update the allocation site target generation
for allocation. This task, however, can be time (and CPU) consuming as it depends
on the number of profiled allocation contexts, and therefore needs to be executed
with minimal interference with regards to the application workload.

Ideally, this task would take place right after a garbage collection cycle, still inside
the safepoint.\footnote{A safepoint is the mechanism used in HotSpot to create
Stop-the-World pauses. Garbage collection cycles run insithe a safepoint, during
which all application threads are stopped.} This would ensure that, when
application threads resume after the safepoint, new target generations for
allocation sites would be selected.

However, executing Alg. \ref{alg:update_tgen} inside a safepoint would
increase its duration, and therefore, would increase application pause times. To
avoid that, the execution of Alg. \ref{alg:update_tgen} is pushed out of
the safepoint, running right after the garbage collection cycle finishes. In
practice, target generation updates are performed concurrently with the application
threads. This leads to a small delay in the adjustment of the allocation site target
generation but also reduces the amount of work (and therefore the duration) of
application pauses (which is our primary goal).

\section{Evaluation}
\label{sec:eval}
This section provides evaluation results for \XPTO. The goal of this evaluation is to show that:
i) profiling information can be efficiently produced and maintained with negligible
throughput and memory overhead;
ii) application pauses can be greatly reduced with \XPTO~when compared to G1;
iii) application pauses are very similar to what is possible to achieve with
NG2C (which requires off-line profiling and programmer knowledge/effort).

G1 is the current default collector for the most recent versions of the
OpenJDK HotSpot JVM;
NG2C is a pretenuring N-Generational GC, which requires off-line profiling and
programmer effort to give hints to the collector about which allocation
sites should be pretenured. CMS, another popular OpenJDK GC, is not evaluated
as it is a throughput oriented GC and thus, presents higher long tail latencies.
Results comparing the performance of CMS with G1 and NG2C can be found the
literature \cite{Rodrigo:2017:arxiv}.

In this evaluation, NG2C represents the best-case scenario for human-provided
hints (since the programmer controls which objects go into which generation);
obviously, this allows more precise pretenuring when compared to \XPTO, which employs
automatic online profiling.

We use as benchmarks three common large-scale production platforms to conduct experiments:
i) Apache Cassandra 2.1.8 \cite{Lakshman:2010}, a large-scale Key-Value store,
ii) Apache Lucene 6.1.0 \cite{Mccandless:2010}, a high performance text search engine,
and iii) GraphChi 0.2.2 \cite{Kyrola:2012}, a large-scale graph computation engine.
A complete description of each workload is presented in Section \ref{sec:workloads}.

\begin{table*}[t]
\centering
\begin{tabular}{ l | r | r | r | r | r | r | r | r | r | r | r | r | r | r}
Workload & LOC & AS & MC & Coll & PAS & PMC & NG2C & SZ (MB) & SA (sec) & Gen 1 & Gen 2 & Gen 3\\
\hline
Cassandra-WI & 195 101 & 3 609 & 20 885 & 4.21 \% & 84  & 408 & 22 & 66KB & 235 & 91 & 12 & 0\\  
Cassandra-RW & 195 101 & 3 609 & 20 885 & 4.21 \% &109 & 480 & 22 & 92KB & 235 & 101 & 19 & 0\\  
Cassandra-RI & 195 101 & 3 609 & 20 885 & 4.21 \%& 107 & 529 & 22 & 30KB & 235 & 11  & 31 & 19\\  
Lucene       &  89 453 & 1 874 &  8 618 & 1.44 \%&26  & 117 & 8 & 16KB & 33   & 10  & 1  & 0\\   
GraphChi-CC  &  18 537 & 2 823 & 12 602 & 0.92 \%& 65  & 56 & 9 & 8KB & 90    & 9   & 1  & 0\\    
GraphChi-PR  &  18 537 & 2 823 & 12 602 & 0.92 \%& 59  & 52 & 9 & 7KB & 90    & 5   & 1  & 0\\    
\end{tabular}
\caption{\XPTO~Profiling Summary}
\label{tab:alloc_sites}
\end{table*}

\subsection{Evaluation Setup}
The evaluation was performed using a server equipped with an Intel Xeon E5505,
with 16 GB of RAM. The server runs Linux 3.13.
Each experiment runs in complete isolation for 5 times (enough to be able
to detect outliers). All workloads run for 30 minutes each. When running each experiment, the first five minutes of execution are
discarded to ensure minimal interference from JVM loading, JIT compilation, etc.

Heap sizes are always fixed. The maximum heap size is set to 12GB while the young
generation size is set to 2GB. According to our experience, these values are
enough to hold the workings set in memory and to avoid premature massive promotion
of objects to older generations (in the case of G1).

W.r.t \XPTO's launch time variables, the length of the object lifetime arrays
(see Figure \ref{fig:global-alloc-counter}) is defined as 16, which is the max object
age in HotSpot. INC\_GEN\_THRES\_FREQ is 4 for all workloads, meaning that profiling
information is analyzed once every 4 collection cycles. EXPAND\_CXT (controls when to
expand a context, see Section \ref{sec:efficient_table}) and INC\_GEN\_THRES (controls when
the target generation is updated, see Section \ref{sec:aop}) are 0.4 and 0.6
respectively for Lucene and GraphChi, and 0.4 and 0.85 for Cassandra. Note that Cassandra
requires a higher INC\_GEN\_THRES to better filter allocation contexts that should
go into other generations.

Finally, as described in Section \ref{sec:static_analysis}, \XPTO's static analyzer can
receive as input, the packages to process. This is specially important for Cassandra
that has a considerable code base size. Hence, the static analyzer received
\texttt{org. apache.cassandra\{db,utils.memory\}} for Cassandra;
\texttt{org. apache.lucene.store} for Lucene; and \texttt{edu.cmu.grapchi. \{datablocks,engine\}} for GraphChi. These
specific packages were selected because they are the ones that deal with most data
in each platofrm. Note that all subpackages are also statically analyzed.

In sum, in order to use \XPTO, the programmer only needs to define three launch time variables,
and select which application packages manage data inside the application. According to our experience,
adapting the launch time variables form the default values and selecting application packages that manage
data if far less time consuming that offline profiling, where multiple workloads must be run from start
to end to build profiling data.

\subsection{Workload Description}
\label{sec:workloads}
This section provides a more complete description of the workloads used
to evaluate \XPTO.

\subsubsection{Cassandra}
\label{sec:cass_workload}
Cassandra runs under 3 different workloads:
i) write intensive workload (2500 read queries and 7500 write queries per second);
iii) read-write workload (5000 read queries and 5000 write queries per second);
iv) read intensive workload (7500 read queries and 2500 write queries per second).
All workloads use the YCSB benchmark tool, a synthetic workload generator which
mirrors real-world settings.\footnote{The Yahoo! Cloud Serving Benchmark (YCSB) is an
open-source benchmarking tool often used to compare NoSQL database systems.}

\subsubsection{Lucene}
\label{sec:lucene_workload}
Lucene is used to build an in-memory text index using a Wikipedia dump
from 2012.\footnote{Wikipedia dumps are available at dumps.wikimedia.org} The
dump has 31GB and is divided in 33M documents. Each document is loaded
into Lucene and can be searched.
The workload is composed by 20000 writes (document updates)
and 5000 reads (document searches) per second; note that this is a
write intensive workload
which represents a worst case scenario for GC pauses.
For reads (document queries), the 500 top words in the dataset are searched in loop;
this also represents a worst case scenario for GC pauses.

\subsubsection{GraphChi}
\label{sec:graphchi_workload}
When compared to the previous systems (Cassandra and Lucene), GraphChi is a
more throughput oriented system (and not latency oriented).
However, GraphChi is used for two reasons:
i) to demonstrate that \XPTO~does not significantly decrease throughput
even in a throughput oriented system;
ii) to demonstrate that, with \XPTO, systems such as GraphChi can now be used
for applications providing latency oriented services, besides performing throughput
oriented graph computations.
In our evaluation, two well-known algorithms are used: i) page rank, and ii) connected
components. Both algorithms use as input a 2010 twitter graph \cite{Kwak:2010}
consisting of 42 millions vertexes and 1.5 billions edges.
These vertexes (and the corresponding edges) are loaded in batches into memory; GraphChi
calculates a memory budget to determine the number of edges to load into
memory before the next batch. This represents an iterative process; in each iteration
a new batch of vertexes is loaded and processed.

\subsection{Application Profiling}
\label{sec:eval_profiling}
This section summarizes the amount of profiling used when evaluating \XPTO~and also compares
it to the amount of human-made code modifications in NG2C. Table \ref{tab:alloc_sites}
presents a number of metrics for each workload:
\textbf{LOC}, lines of code of the platform;
\textbf{AS}/\textbf{MC}, number of allocation sites / method calls suggested for profiling by the bytecode static analyzer;
\textbf{Coll}, number of allocation sites with more than one call graph path producing the same context summary (collision);
\textbf{PAS}/\textbf{PMC}, number of profiled allocation sites / method calls (i.e., allocation sites / method calls where profiling code was actually inserted);
\textbf{NG2C}, number of code locations that were changed to evaluate NG2C (as previously reported \cite{Rodrigo:2017:arxiv});
\textbf{SZ}, memory overhead of the object lifetime distribution table (see Figure \ref{fig:global-alloc-counter});
\textbf{SA}, the amount of time required to run the static analyzer;
\textbf{Gen X}, number of allocation contexts in a specific generation.

From Table \ref{tab:alloc_sites}, five important points must be retained.
Fist, looking at PAS and PMC, the number of hot/jitted allocation sites
and method calls is small.
This demonstrates
that the profiling effort is greatly reduced by only profiling hot code locations.
Second, looking at SZ, the memory overhead introduced to support profiling information
does not exceed 92KB, a reasonable memory overhead considering the performance advantages
that can be achieved by leveraging the information in it.
Third, looking at SA, the time necessary to run the static analyzer does not exceed 4 minutes,
a manageable cost for a task that can be done off-line (and therefore can be
amortized through numerous executions of applications).
Fourth, looking at Gen X columns, only three generations were used to
partition allocation contexts (i.e., no allocation context was moved to Gen 4 and above).
Finally, the percentage of allocation sites with collisions (with more than one call graph path
producing the same context summary) does not exceed 4.21\%, showing that, despite using a weak hash
construction (based on addition and subtraction of hashes), it is possible to achieve a low collision
rate compared to previous works (see Section \ref{sec:rel_work}). This mostly comes from the fact that only hot methods are profiled, thus leading to a reduction
in the actual number of collisions.

It is also interesting to note that all the code changes used in NG2C (which require
off-line profiling and human knowledge) are automatically identified and profiled in
\XPTO. \XPTO~additionally profiles other code locations (which are not used for NG2C),
leading to additional improvements.

\subsection{Pause Time Percentiles and Distribution}
\begin{figure*}[!t]
    \centering
    \begin{subfigure}{.32\textwidth}
        \includegraphics[width=\textwidth]{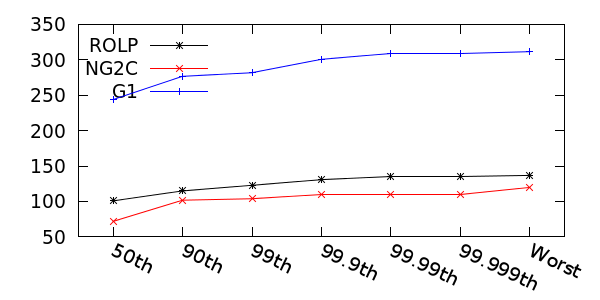}
        \caption{Cassandra WI}
    \end{subfigure}
    \begin{subfigure}{.32\textwidth}
        \includegraphics[width=\textwidth]{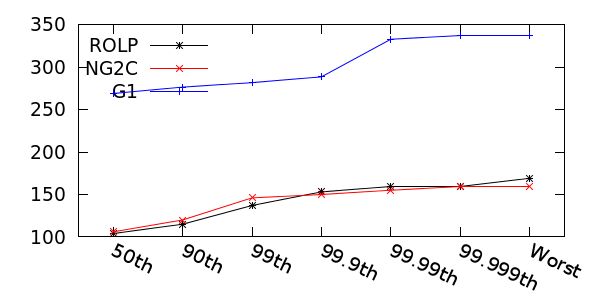}
        \caption{Cassandra WR}
    \end{subfigure}
    \begin{subfigure}{.32\textwidth}
        \includegraphics[width=\textwidth]{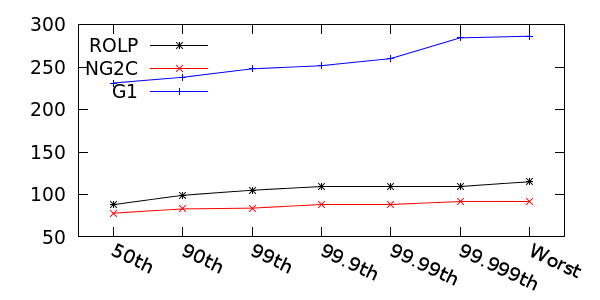}
        \caption{Cassandra RI}
    \end{subfigure}
    \begin{subfigure}{.32\textwidth}
        \includegraphics[width=\textwidth]{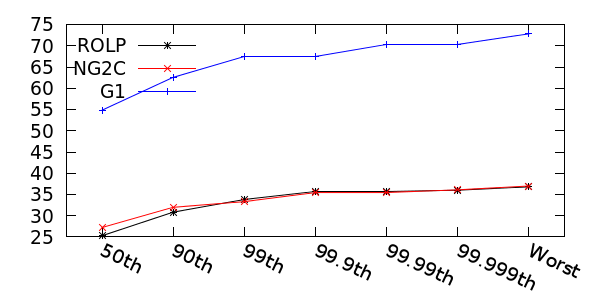}
        \caption{Lucene}
    \end{subfigure}
    \begin{subfigure}{.32\textwidth}
        \includegraphics[width=\textwidth]{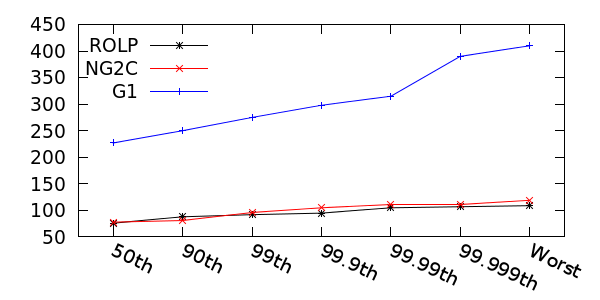}
        \caption{GraphChi CC}
    \end{subfigure}
    \begin{subfigure}{.32\textwidth}
        \includegraphics[width=\textwidth]{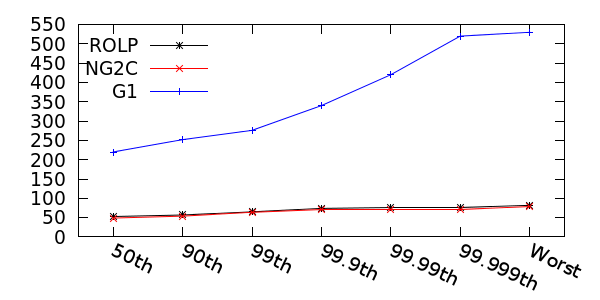}
        \caption{GraphChi PR}
    \end{subfigure}
\caption{Pause Time Percentiles (ms)}
\label{fig:pause_percentiles}
\end{figure*}

Figure \ref{fig:pause_percentiles} presents the results for application
pauses across all workloads, for \XPTO, NG2C, and G1. Pauses are presented in
milliseconds and are organized by percentiles.

Compared to G1, \XPTO~significantly improves application pauses for all percentiles across all
workloads. Regarding NG2C (which requires developer knowledge), \XPTO~approaches
the numbers provided by NG2C for most workloads. Only for Cassandra workloads,
there is a very small performance difference between \XPTO~and NG2C. This comes
from the fact that Cassandra is a very complex platform and sometimes it takes
time for \XPTO~to find the appropriate generation for each allocation context.

From these results, the main conclusion to take is that \XPTO~can significantly
reduce long tail latencies when compared to G1, the most advanced GC implementation in OpenJDK HotSpot;
in addition, it can also keep up with NG2C, but without requiring
any programming effort and knowledge.

\begin{figure*}[!t]
\centering
    \begin{subfigure}{.32\textwidth}
        \caption{Cassandra WI}
        \includegraphics[width=\textwidth]{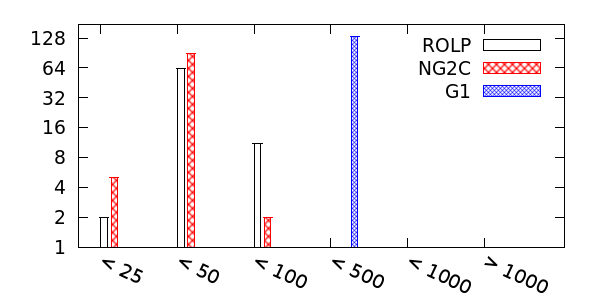}
    \end{subfigure}
    \begin{subfigure}{.32\textwidth}
        \caption{Cassandra WR}
        \includegraphics[width=\textwidth]{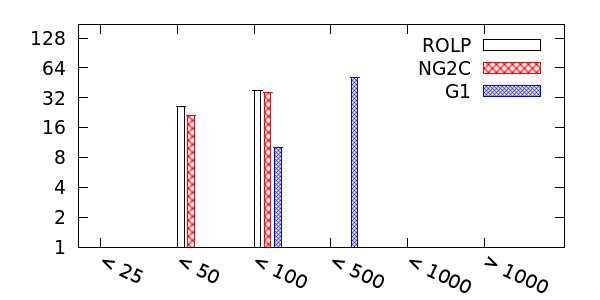}
    \end{subfigure}
    \begin{subfigure}{.32\textwidth}
        \caption{Cassandra RI}
        \includegraphics[width=\textwidth]{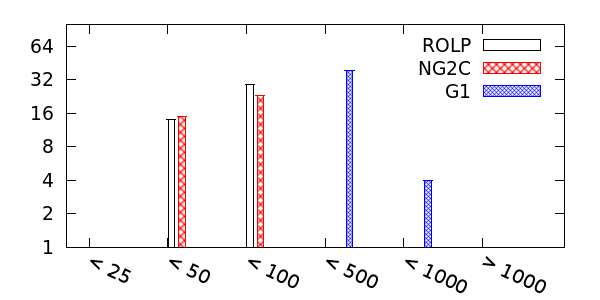}
    \end{subfigure}
    \begin{subfigure}{.32\textwidth}
        \caption{Lucene}
        \includegraphics[width=\textwidth]{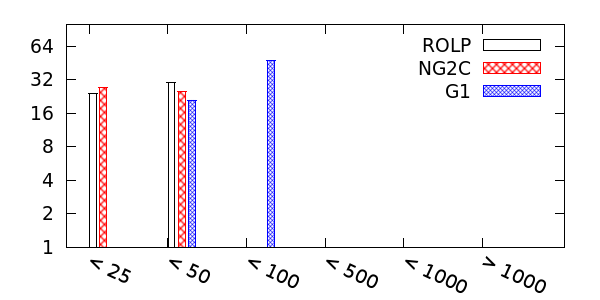}
    \end{subfigure}
    \begin{subfigure}{.32\textwidth}
        \caption{GraphChi CC}
        \includegraphics[width=\textwidth]{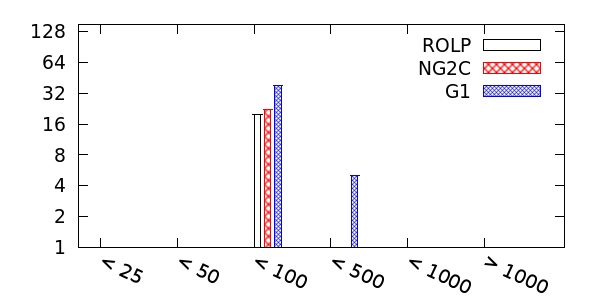}
    \end{subfigure}
    \begin{subfigure}{.32\textwidth}
        \caption{GraphChi PR}
        \includegraphics[width=\textwidth]{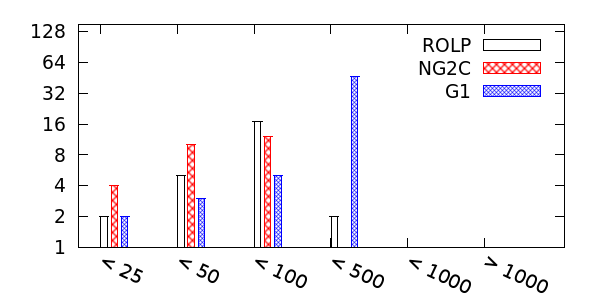}
    \end{subfigure}
\caption{Number of Application Pauses Per Duration Interval (ms)}
\label{fig:pause_distribution}
\end{figure*}

So far, the presented application pause times were organized by percentiles.
Figure \ref{fig:pause_distribution}
presents the number of application pauses that occur in each pause time interval.
Pauses with shorter durations appear in intervals to the left while longer pauses
appear in intervals to the right. In other words, the less pauses to the right,
the better.

\XPTO~presents significant improvements regarding G1, i.e., it results in less
application pauses in longer intervals, across all workloads. When comparing
\XPTO~with NG2C, both solutions present very similar pause time distribution.

In sum, \XPTO~reduces application pauses by automatically pretenuring objects
from allocation contexts that tend to allocate objects with longer lifetimes. When compared
to G1, \XPTO~greatly reduces application pauses and object copying within the heap.
Compared to NG2C, \XPTO~presents equivalent performance without
requiring programmer effort and knowledge.

\subsection{Throughput and Memory Usage}\begin{figure}[!t]
\centering
    \begin{subfigure}{.32\textwidth}
        \caption{Throughput}
        \includegraphics[width=\textwidth]{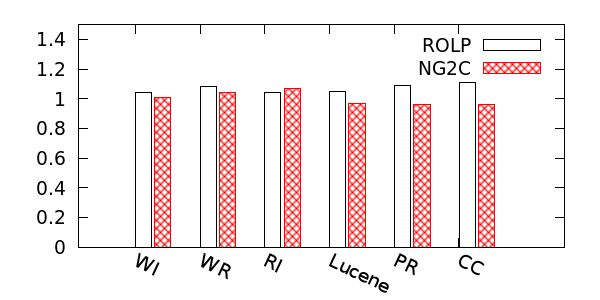}
    \end{subfigure}
    \begin{subfigure}{.32\textwidth}
        \caption{Max Memory Usage}
        \includegraphics[width=\textwidth]{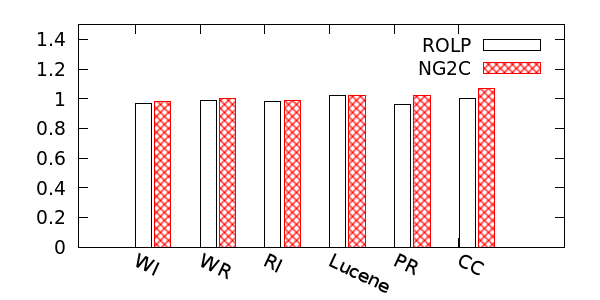}
    \end{subfigure}
\caption{Throughput and Max Memory Usage norm. to G1}
\label{fig:throughput_memory}
\end{figure}


This section shows results on application throughput and max memory usage for G1, NG2C, and
\XPTO. The goal of this section is to demonstrate that: i) \XPTO~does not inflict a significant throughput overhead due to its profiling code, and ii) \XPTO~does not
negatively impact the max memory usage.

Figure \ref{fig:throughput_memory} shows results for the throughput and max memory
usage. All results are normalized to G1 (i.e., all G1 would be plotted as 1 for all
columns). \XPTO~presents a negligible throughput
decrease, less than 5\% (on average) for most workloads, compared to G1. Only for GraphChi workloads,
\XPTO~presents an average throughput overhead above 5\% (9\% for PR and 11\% for CC). This
is still a manageable throughput overhead (eg., 198 seconds in 30 min of execution with 11\% overhead)  considering the great reduction in application
long tail latencies.

Note that this throughput overhead could be removed by dynamically removing profiling code from
allocation sites which allocate very short lived objects. \XPTO~currently does not
implement this optimization; however, it is being considered for future work.
Since NG2C does not employ any online profiling, there is no throughput overhead. Regarding
max heap usage, both \XPTO~and NG2C do not present any relevant overhead or improvement.
Essentially, the max memory usage is not affected.


\section{Related Work} 
\label{sec:rel_work}
Profiling plays a key role in managed runtimes, either for code optimization or memory management decisions \cite{arnold:2005,Arnold:2000,Harris:2000,Hertz:2006,Ricci:2011,Xu:2013,Zheng:2015}.
We focus on getting quality profiling information to drive object pretenuring.
\XPTO~is, to the best of our knowledge, the first online profiler targeting the dynamic pretenuring of objects in Big Data applications running
on HotSpot.
This section compares our work with state-of-art systems, namely, off-line and online profilers that guide systems where small changes are needed in the heap organization and collection. It ends with a comparative analysis of systems that demand a more profound change, either to the application/framework or the runtime itself, in some cases manipulating application-defined types and/or organizing the heap in special purpose regions, and placing data directly in an off-heap space.

\subsection{Off-line Profiling}
Hertz et al. \cite{Hertz:2006} introduced an algorithm where an object lifetime is tracked based on timestamps assigned when the object lost an incoming reference, and when GC identifies it as an unreachable object. This is implemented in a tracing tool called Merlin that, when is analyzing a program, it can be up to 300 times slower compared to a non-profiled run. Ricci et al. \cite{Ricci:2011} uses the same algorithm but adds new functionalities to his Elephant track tool in terms of precision and comprehensiveness of reference sources (i.e.\ weak reference). Another system, Resurrector \cite{Xu:2013}, relaxes on precision to provide faster profiling but  still introduces 3 to 40 times slowdown depending on the workload.

Blackburn et al. \cite{Blackburn:2007} extends the profile-based pretenuring of Cheng's solution \cite{Cheng:1998} using the Merlin tracing tool \cite{Hertz:2006}. They have a two stage profiling. The first stage happens during the build of the JVM\@. Profiling at this stage is used to improve the performance of JVM itself, since Jikes RVM \cite{Arnold:2000} is a meta-circular JVM.
Blackburn et al. reports this is particularly useful for tight heaps (i.e.\ heaps that are just above the minimum size for a given application, reaching at most 150MB) and not suitable for heaps with Gigabytes of objects. The second stage is an application-specific process, based on the off-line profiling made with Merlin~\cite{Hertz:2006}.

Sewe et al.~\cite{Sewe:2010} presents an headroom schema which drives pretenuring based on the space left on the heap before garbage collection is necessary.
Although their solution brings advantages to collection times, they push much of the overhead to the  mutator and also to the off-line process, which is not always possible or accurate. This approach makes the classification not only dependent on the application but also on the overall heap size. Finally, Sewe et al.\cite{Sewe:2010} do not target large heaps or a modern garbage collector like G1.

NG2C~\cite{Rodrigo:2017:arxiv} extends G1 to support object pretenuring. However, it also needs offline profiling and programmer's help to identify the generation where a new object should be allocated. Thus, we can say that it uses an off-line profiler to establish a relation between allocation sites and object lifetimes, missing the opportunity to avoid inhomogeneous allocation behavior~\cite{Jones:2008}.
Cohen et al.~\cite{data-ismm-15} extends the operation of the Immix garbage collector in Jikes RVM \cite{Blackburn:2008} with a new programming interface between the application and the GC, in order to manage dominant data structures (i.e. a data structure holding most of the objects during the lifetime of the program) more efficiently.
The main advantage comes from reducing the occurrence of highly entangled deep-shaped data structures lay-out in memory, thus improving performance of the parallel tracing stage.

Compared to previous solutions, \XPTO~does not require any source code modifications, it targets a widely employed industrial Java VM, and was tested using large heap sizes.


\subsection{Online Profiling}

The previous profilers produce traces that are then used for posteriori optimizations, either manual or automated. In general, input influences the choices made during memory management \cite{Mao:2009} so, profiling online can take into account fresh data coming from a new execution.
\XPTO~relies on a class of profiling techniques known as feedback-directed optimization (FDO), as described originally by Smith \cite{Smith:2000}.
However, memory organization decisions based on online profiling can impose a significant overhead in collecting and processing information to apply transformations.

FDO techniques can be organized into four categories~\cite{arnold:2005}: i)  runtime service monitoring, ii) hardware performance monitors, iii) sampling, and iv) program instrumentation.
In \XPTO~we use a mix of runtime monitoring, with some data already collected by services of the runtime, and lightweight program instrumentation to collect extra information and make pretenuring decisions.

\XPTO~ needs the {\it calling context} to profile objects at relevant allocation sites.
Ball and Laurus \cite{Ball:1996} compute a unique number for each possible path of a control flow graph inside a procedure. The computation is done offline and added to the source code. This is not suited for \XPTO~ because modern workloads have many possible paths inside each routine, and the technique can not capture the inter-procedure path needed for \XPTO~ to distinguish allocation sites.
Bond and McKinley \cite{Bond:2007} also compute a single value, but at runtime, to determine the sequence of active stack frames in a inter-procedural way. However, they need to maintain non-commutativity to differentiate call orders, such as calling methods $A\rightarrow B \rightarrow C$ and $B\rightarrow A \rightarrow C$. This is not a requirement for \XPTO~ and so we can have a smaller impact on code instrumentation, with low impact on the throughput. \XPTO~ uses an adaptive solution to store allocation contexts, with a low ratio of collisions (as shown in Table ~\ref{tab:alloc_sites}) while using fewer bits to store information.



%





Clifford et al.~\cite{folding-ismm-14} performs online profiling to optimize memory allocation operations; it performs a single allocation (improvement over allocation inlining) of a block big enough to fit multiple objects that are allocated in sequence. Objects need not have parent-child or sibling relationships among them because it is the control-flow relations between hot code flows and locations that are monitored, as we address in our work. So, while allocations may take place in different methods, equivalent GC behavior is ensured. However, while doing online profiling, it only addresses initial allocation and not the issue of pretenuring objects.

Memento~\cite{Clifford:2015} gathers online temporal feedback regarding object lifetime by instrumenting allocation and leaving \emph{mementos} alongside objects in the nursery, so that the allocation site of the object can be recorded. When objects are eventually tenured, Memento is able to avoid later the overhead of scanning and copying for objects allocated in the same allocation site. As in our work, it attempts to allocate these objects directly in the tenure space, as they will be long lived; however, it is only able to manage one tenured space, therefore applying a binary decision that will still potentially co-locate objects with possibly very different lifetimes, incurring in additional compaction effort. Our work manages multiple spaces  and is therefore able to allocate objects directly in the space where objects with similar lifetime will also reside. In addition, Memento instruments all application code
while it is still being interpreted. This has two disadvantages compared to \XPTO: i) all the
application code is being profiled, leading to a huge profiling overhead (in \XPTO, we only
profile hot code locations); ii) profiling stops when the code is JIT compiled, meaning
that application behavior is only tracked while the application is starting and the code
is not jitted. This represents a problem if the application has a long startup time or
if workloads change since the profiling decisions are taken and cannot be changed.
Finally, Memento does not track allocation contexts (i.e. call graph), which we found
to be essential to properly profile complex platforms such as Cassandra.

\subsection{Big Data Garbage Collectors}

Others systems 
employ a less transparent approach (from the source code perspective) to handle large heaps while taking into account the organization of typical big data applications.
However, these systems either depend on large modifications to the heap organization and collection and/or an important collaboration of the programmer to mark parts of the code (in some cases the framework) to be treated specially.

Facade ~\cite{Nguyen:2015} is a compiler --- an extension to Soot~\cite{Vallee-Rai:2000} --- which reduces the number of objects in the heap by separating data (fields) from control (methods) and putting data objects in an off-heap structure without the need to maintain the bloat-causing header \cite{Bu:2013}.
However, the programmer is responsible by a time consuming task, which has to be repeated for each new application or framework: to identify the control path and the list of classes that make up the data path.
A similar approach is followed by Broom~\cite{Gog:2015} where the heap is split into regions \cite{Tofte:1997} explicitly created by the programmer (assumed to know which codebase creates related objects).
Yak~\cite{Nguyen:2016} minimizes this effort but still relies on the programmer to identify epochs, which is another way to identify objects that have a similar lifetime. These objects can be allocated in the same region, avoiding a full heap / generation tracing to identify dead objects. However, it requires not only the programmer to have access to the source code and understand where to place the limits of each region,
but also new bookkeeping structures for inter-reference spaces must be put in place.

Sources of overhead in Big Data applications can be found not only on poor object placement or the excessive bloat produced by an objects' header, but also by garbage collections running uncoordinated in inter-dependent JVM instances.
When a group of JVMs are coordinated in a distributed task and need to collect unreachable objects, if they do so regardless of each other, this can cause several instances to wait for each other resulting in significant pause times \cite{Maas:2016}. Our approach can complement these systems and, because we only use online information, \XPTO~does not need to synchronize off-line profiling information across a cluster of JVMs.

\section{Conclusions}
\label{sec:conclusions}
This paper presents \XPTO, a runtime allocation site profiler that helps
generational garbage collectors to decide whether and where to pretenure objects.
\XPTO~is implemented for the OpenJDK 8 HotSpot, one of the most used JVM implementations. Although
\XPTO~is generic and could be used with other generational collectors, for this work,
we used NG2C, a pretenuring N-Generational collector also available for HotSpot.
\XPTO~uses efficient profiling code that allows the identification of allocation
sites which allocate objects with long lifetimes. By combining this profiling
information with NG2C, it is possible to reduce application pauses with no significant
throughput overhead.
\XPTO~represents a drop-in replacement for the
HotSpot JVM which, with no programmer effort, reduces application pause times.
To the best of our knowledge, this is the first solution to propose online profiling
for reducing application pause times in HotSpot while running production workloads. \XPTO~is
open source and it is available at github.com/paper-168/rolp.

\bibliography{main}


\begin{thebibliography}{38}


\ifx \showCODEN    \undefined \def \showCODEN     #1{\unskip}     \fi
\ifx \showDOI      \undefined \def \showDOI       #1{#1}\fi
\ifx \showISBNx    \undefined \def \showISBNx     #1{\unskip}     \fi
\ifx \showISBNxiii \undefined \def \showISBNxiii  #1{\unskip}     \fi
\ifx \showISSN     \undefined \def \showISSN      #1{\unskip}     \fi
\ifx \showLCCN     \undefined \def \showLCCN      #1{\unskip}     \fi
\ifx \shownote     \undefined \def \shownote      #1{#1}          \fi
\ifx \showarticletitle \undefined \def \showarticletitle #1{#1}   \fi
\ifx \showURL      \undefined \def \showURL       {\relax}        \fi
\providecommand\bibfield[2]{#2}
\providecommand\bibinfo[2]{#2}
\providecommand\natexlab[1]{#1}
\providecommand\showeprint[2][]{arXiv:#2}

\bibitem[\protect\citeauthoryear{Arnold, Fink, Grove, Hind, and Sweeney}{Arnold
  et~al\mbox{.}}{2000}]%
        {Arnold:2000}
\bibfield{author}{\bibinfo{person}{Matthew Arnold}, \bibinfo{person}{Stephen
  Fink}, \bibinfo{person}{David Grove}, \bibinfo{person}{Michael Hind}, {and}
  \bibinfo{person}{Peter~F. Sweeney}.} \bibinfo{year}{2000}\natexlab{}.
\newblock \showarticletitle{Adaptive Optimization in the Jalape\~{N}O JVM}. In
  \bibinfo{booktitle}{{\em Proceedings of the 15th ACM SIGPLAN Conference on
  Object-oriented Programming, Systems, Languages, and Applications}} {\em
  (\bibinfo{series}{OOPSLA '00})}. \bibinfo{publisher}{ACM},
  \bibinfo{address}{New York, NY, USA}, \bibinfo{pages}{47--65}.
\newblock
\showISBNx{1-58113-200-X}
\showDOI{%
\url{https://doi.org/10.1145/353171.353175}}


\bibitem[\protect\citeauthoryear{Arnold, Fink, Grove, Hind, and Sweeney}{Arnold
  et~al\mbox{.}}{2005}]%
        {arnold:2005}
\bibfield{author}{\bibinfo{person}{M. Arnold}, \bibinfo{person}{S.~J. Fink},
  \bibinfo{person}{D. Grove}, \bibinfo{person}{M. Hind}, {and}
  \bibinfo{person}{P.~F. Sweeney}.} \bibinfo{year}{2005}\natexlab{}.
\newblock \showarticletitle{A Survey of Adaptive Optimization in Virtual
  Machines}.
\newblock \bibinfo{journal}{{\it Proc. IEEE}} \bibinfo{volume}{93},
  \bibinfo{number}{2} (\bibinfo{date}{Feb} \bibinfo{year}{2005}),
  \bibinfo{pages}{449--466}.
\newblock
\showISSN{0018-9219}


\bibitem[\protect\citeauthoryear{Ball and Larus}{Ball and Larus}{1996}]%
        {Ball:1996}
\bibfield{author}{\bibinfo{person}{Thomas Ball} {and} \bibinfo{person}{James~R.
  Larus}.} \bibinfo{year}{1996}\natexlab{}.
\newblock \showarticletitle{Efficient Path Profiling}. In
  \bibinfo{booktitle}{{\em Proceedings of the 29th Annual ACM/IEEE
  International Symposium on Microarchitecture}} {\em (\bibinfo{series}{MICRO
  29})}. \bibinfo{publisher}{IEEE Computer Society},
  \bibinfo{address}{Washington, DC, USA}, \bibinfo{pages}{46--57}.
\newblock
\showISBNx{0-8186-7641-8}
\showURL{%
\url{http://dl.acm.org/citation.cfm?id=243846.243857}}


\bibitem[\protect\citeauthoryear{Blackburn, Hertz, Mckinley, Moss, and
  Yang}{Blackburn et~al\mbox{.}}{2007}]%
        {Blackburn:2007}
\bibfield{author}{\bibinfo{person}{Stephen~M. Blackburn},
  \bibinfo{person}{Matthew Hertz}, \bibinfo{person}{Kathryn~S. Mckinley},
  \bibinfo{person}{J.~Eliot~B. Moss}, {and} \bibinfo{person}{Ting Yang}.}
  \bibinfo{year}{2007}\natexlab{}.
\newblock \showarticletitle{Profile-based Pretenuring}.
\newblock \bibinfo{journal}{{\em ACM Trans. Program. Lang. Syst.\/}}
  \bibinfo{volume}{29}, \bibinfo{number}{1}, Article \bibinfo{articleno}{2}
  (\bibinfo{date}{Jan.} \bibinfo{year}{2007}).
\newblock
\showISSN{0164-0925}
\showURL{%
\url{http://doi.acm.org/10.1145/1180475.1180477}}


\bibitem[\protect\citeauthoryear{Blackburn and McKinley}{Blackburn and
  McKinley}{2008}]%
        {Blackburn:2008}
\bibfield{author}{\bibinfo{person}{Stephen~M. Blackburn} {and}
  \bibinfo{person}{Kathryn~S. McKinley}.} \bibinfo{year}{2008}\natexlab{}.
\newblock \showarticletitle{Immix: A Mark-region Garbage Collector with Space
  Efficiency, Fast Collection, and Mutator Performance}. In
  \bibinfo{booktitle}{{\em Proceedings of the 29th ACM SIGPLAN Conference on
  Programming Language Design and Implementation}} {\em (\bibinfo{series}{PLDI
  '08})}. \bibinfo{publisher}{ACM}, \bibinfo{pages}{22--32}.
\newblock
\showISBNx{978-1-59593-860-2}
\showURL{%
\url{http://doi.acm.org/10.1145/1375581.1375586}}


\bibitem[\protect\citeauthoryear{Bond and McKinley}{Bond and McKinley}{2007}]%
        {Bond:2007}
\bibfield{author}{\bibinfo{person}{Michael~D. Bond} {and}
  \bibinfo{person}{Kathryn~S. McKinley}.} \bibinfo{year}{2007}\natexlab{}.
\newblock \showarticletitle{Probabilistic Calling Context}. In
  \bibinfo{booktitle}{{\em Proceedings of the 22Nd Annual ACM SIGPLAN
  Conference on Object-oriented Programming Systems and Applications}} {\em
  (\bibinfo{series}{OOPSLA '07})}. \bibinfo{publisher}{ACM},
  \bibinfo{address}{New York, NY, USA}, \bibinfo{pages}{97--112}.
\newblock
\showISBNx{978-1-59593-786-5}
\showDOI{%
\url{https://doi.org/10.1145/1297027.1297035}}


\bibitem[\protect\citeauthoryear{Bruno, Oliveira, and Ferreira}{Bruno
  et~al\mbox{.}}{2017}]%
        {Rodrigo:2017:arxiv}
\bibfield{author}{\bibinfo{person}{Rodrigo Bruno},
  \bibinfo{person}{Lu\'{\i}s~Picciochi Oliveira}, {and} \bibinfo{person}{Paulo
  Ferreira}.} \bibinfo{year}{2017}\natexlab{}.
\newblock \showarticletitle{NG2C: Pretenuring Garbage Collection with Dynamic
  Generations for HotSpot Big Data Applications}. In \bibinfo{booktitle}{{\em
  Proceedings of the 2017 ACM SIGPLAN International Symposium on Memory
  Management}} {\em (\bibinfo{series}{ISMM 2017})}. \bibinfo{publisher}{ACM},
  \bibinfo{address}{New York, NY, USA}, \bibinfo{pages}{2--13}.
\newblock
\showISBNx{978-1-4503-5044-0}
\showDOI{%
\url{https://doi.org/10.1145/3092255.3092272}}


\bibitem[\protect\citeauthoryear{Bu, Borkar, Xu, and Carey}{Bu
  et~al\mbox{.}}{2013}]%
        {Bu:2013}
\bibfield{author}{\bibinfo{person}{Yingyi Bu}, \bibinfo{person}{Vinayak
  Borkar}, \bibinfo{person}{Guoqing Xu}, {and} \bibinfo{person}{Michael~J.
  Carey}.} \bibinfo{year}{2013}\natexlab{}.
\newblock \showarticletitle{A Bloat-aware Design for Big Data Applications}. In
  \bibinfo{booktitle}{{\em Proceedings of the 2013 International Symposium on
  Memory Management}} {\em (\bibinfo{series}{ISMM '13})}.
  \bibinfo{publisher}{ACM}, \bibinfo{address}{New York, NY, USA},
  \bibinfo{pages}{119--130}.
\newblock
\showISBNx{978-1-4503-2100-6}
\showDOI{%
\url{https://doi.org/10.1145/2464157.2466485}}


\bibitem[\protect\citeauthoryear{Cheng, Harper, and Lee}{Cheng
  et~al\mbox{.}}{1998}]%
        {Cheng:1998}
\bibfield{author}{\bibinfo{person}{Perry Cheng}, \bibinfo{person}{Robert
  Harper}, {and} \bibinfo{person}{Peter Lee}.} \bibinfo{year}{1998}\natexlab{}.
\newblock \showarticletitle{Generational Stack Collection and Profile-driven
  Pretenuring}. In \bibinfo{booktitle}{{\em Proceedings of the ACM SIGPLAN 1998
  Conference on Programming Language Design and Implementation}} {\em
  (\bibinfo{series}{PLDI '98})}. \bibinfo{pages}{162--173}.
\newblock
\showISBNx{0-89791-987-4}
\showURL{%
\url{http://doi.acm.org/10.1145/277650.277718}}


\bibitem[\protect\citeauthoryear{Clifford, Payer, Stanton, and Titzer}{Clifford
  et~al\mbox{.}}{2015}]%
        {Clifford:2015}
\bibfield{author}{\bibinfo{person}{Daniel Clifford}, \bibinfo{person}{Hannes
  Payer}, \bibinfo{person}{Michael Stanton}, {and} \bibinfo{person}{Ben~L.
  Titzer}.} \bibinfo{year}{2015}\natexlab{}.
\newblock \showarticletitle{Memento Mori: Dynamic Allocation-site-based
  Optimizations}. In \bibinfo{booktitle}{{\em Proceedings of the 2015
  International Symposium on Memory Management}} {\em (\bibinfo{series}{ISMM
  '15})}. \bibinfo{publisher}{ACM}, \bibinfo{address}{New York, NY, USA},
  \bibinfo{pages}{105--117}.
\newblock
\showISBNx{978-1-4503-3589-8}
\showDOI{%
\url{https://doi.org/10.1145/2754169.2754181}}


\bibitem[\protect\citeauthoryear{Clifford, Payer, Starzinger, and
  Titzer}{Clifford et~al\mbox{.}}{2014}]%
        {folding-ismm-14}
\bibfield{author}{\bibinfo{person}{Daniel Clifford}, \bibinfo{person}{Hannes
  Payer}, \bibinfo{person}{Michael Starzinger}, {and} \bibinfo{person}{Ben~L.
  Titzer}.} \bibinfo{year}{2014}\natexlab{}.
\newblock \showarticletitle{Allocation Folding Based on Dominance}. In
  \bibinfo{booktitle}{{\em Proceedings of the 2014 International Symposium on
  Memory Management}} {\em (\bibinfo{series}{ISMM '14})}.
  \bibinfo{publisher}{ACM}, \bibinfo{address}{New York, NY, USA},
  \bibinfo{pages}{15--24}.
\newblock
\showISBNx{978-1-4503-2921-7}
\showDOI{%
\url{https://doi.org/10.1145/2602988.2602994}}


\bibitem[\protect\citeauthoryear{Cohen and Petrank}{Cohen and Petrank}{2015}]%
        {data-ismm-15}
\bibfield{author}{\bibinfo{person}{Nachshon Cohen} {and} \bibinfo{person}{Erez
  Petrank}.} \bibinfo{year}{2015}\natexlab{}.
\newblock \showarticletitle{Data Structure Aware Garbage Collector}. In
  \bibinfo{booktitle}{{\em Proceedings of the 2015 International Symposium on
  Memory Management}} {\em (\bibinfo{series}{ISMM '15})}.
  \bibinfo{publisher}{ACM}, \bibinfo{address}{New York, NY, USA},
  \bibinfo{pages}{28--40}.
\newblock
\showISBNx{978-1-4503-3589-8}
\showDOI{%
\url{https://doi.org/10.1145/2754169.2754176}}


\bibitem[\protect\citeauthoryear{Detlefs, Flood, Heller, and Printezis}{Detlefs
  et~al\mbox{.}}{2004}]%
        {Detlefs:2004}
\bibfield{author}{\bibinfo{person}{David Detlefs}, \bibinfo{person}{Christine
  Flood}, \bibinfo{person}{Steve Heller}, {and} \bibinfo{person}{Tony
  Printezis}.} \bibinfo{year}{2004}\natexlab{}.
\newblock \showarticletitle{Garbage-first Garbage Collection}. In
  \bibinfo{booktitle}{{\em Proceedings of the 4th International Symposium on
  Memory Management}} {\em (\bibinfo{series}{ISMM '04})}.
  \bibinfo{publisher}{ACM}, \bibinfo{address}{New York, NY, USA},
  \bibinfo{pages}{37--48}.
\newblock
\showISBNx{1-58113-945-4}
\showDOI{%
\url{https://doi.org/10.1145/1029873.1029879}}


\bibitem[\protect\citeauthoryear{Dice, Moir, and Scherer}{Dice
  et~al\mbox{.}}{2010}]%
        {dice2010quickly}
\bibfield{author}{\bibinfo{person}{D. Dice}, \bibinfo{person}{M.S. Moir}, {and}
  \bibinfo{person}{W.N. Scherer}.} \bibinfo{year}{2010}\natexlab{}.
\newblock \bibinfo{title}{Quickly reacquirable locks}.
\newblock   (\bibinfo{date}{Oct.~12} \bibinfo{year}{2010}).
\newblock
\showURL{%
\url{https://www.google.ch/patents/US7814488}}
\newblock
\shownote{US Patent 7,814,488.}


\bibitem[\protect\citeauthoryear{Gidra, Thomas, Sopena, and Shapiro}{Gidra
  et~al\mbox{.}}{2012}]%
        {Gidra:2012}
\bibfield{author}{\bibinfo{person}{Lokesh Gidra}, \bibinfo{person}{Ga\"{e}l
  Thomas}, \bibinfo{person}{Julien Sopena}, {and} \bibinfo{person}{Marc
  Shapiro}.} \bibinfo{year}{2012}\natexlab{}.
\newblock \showarticletitle{Assessing the Scalability of Garbage Collectors on
  Many Cores}.
\newblock \bibinfo{journal}{{\em SIGOPS Oper. Syst. Rev.\/}}
  \bibinfo{volume}{45}, \bibinfo{number}{3} (\bibinfo{date}{Jan.}
  \bibinfo{year}{2012}), \bibinfo{pages}{15--19}.
\newblock
\showISSN{0163-5980}
\showDOI{%
\url{https://doi.org/10.1145/2094091.2094096}}


\bibitem[\protect\citeauthoryear{Gidra, Thomas, Sopena, and Shapiro}{Gidra
  et~al\mbox{.}}{2013}]%
        {Gidra:2013}
\bibfield{author}{\bibinfo{person}{Lokesh Gidra}, \bibinfo{person}{Ga\"{e}l
  Thomas}, \bibinfo{person}{Julien Sopena}, {and} \bibinfo{person}{Marc
  Shapiro}.} \bibinfo{year}{2013}\natexlab{}.
\newblock \showarticletitle{A Study of the Scalability of Stop-the-world
  Garbage Collectors on Multicores}. In \bibinfo{booktitle}{{\em Proceedings of
  the Eighteenth International Conference on Architectural Support for
  Programming Languages and Operating Systems}} {\em (\bibinfo{series}{ASPLOS
  '13})}. \bibinfo{publisher}{ACM}, \bibinfo{address}{New York, NY, USA},
  \bibinfo{pages}{229--240}.
\newblock
\showISBNx{978-1-4503-1870-9}
\showDOI{%
\url{https://doi.org/10.1145/2451116.2451142}}


\bibitem[\protect\citeauthoryear{Gog, Giceva, Schwarzkopf, Vaswani, Vytiniotis,
  Ramalingam, Costa, Murray, Hand, and Isard}{Gog et~al\mbox{.}}{2015}]%
        {Gog:2015}
\bibfield{author}{\bibinfo{person}{Ionel Gog}, \bibinfo{person}{Jana Giceva},
  \bibinfo{person}{Malte Schwarzkopf}, \bibinfo{person}{Kapil Vaswani},
  \bibinfo{person}{Dimitrios Vytiniotis}, \bibinfo{person}{Ganesan Ramalingam},
  \bibinfo{person}{Manuel Costa}, \bibinfo{person}{Derek~G. Murray},
  \bibinfo{person}{Steven Hand}, {and} \bibinfo{person}{Michael Isard}.}
  \bibinfo{year}{2015}\natexlab{}.
\newblock \showarticletitle{Broom: Sweeping Out Garbage Collection from Big
  Data Systems}. In \bibinfo{booktitle}{{\em 15th Workshop on Hot Topics in
  Operating Systems (HotOS XV)}}. \bibinfo{publisher}{USENIX Association},
  \bibinfo{address}{Kartause Ittingen, Switzerland}.
\newblock
\showURL{%
\url{https://www.usenix.org/conference/hotos15/workshop-program/presentation/gog}}


\bibitem[\protect\citeauthoryear{Harris}{Harris}{2000}]%
        {Harris:2000}
\bibfield{author}{\bibinfo{person}{Timothy~L. Harris}.}
  \bibinfo{year}{2000}\natexlab{}.
\newblock \showarticletitle{Dynamic Adaptive Pre-tenuring}. In
  \bibinfo{booktitle}{{\em Proceedings of the 2nd International Symposium on
  Memory Management}} {\em (\bibinfo{series}{ISMM '00})}.
  \bibinfo{publisher}{ACM}, \bibinfo{pages}{127--136}.
\newblock
\showISBNx{1-58113-263-8}
\showURL{%
\url{http://doi.acm.org/10.1145/362422.362476}}


\bibitem[\protect\citeauthoryear{Hertz, Blackburn, Moss, McKinley, and
  Stefanovi\'{c}}{Hertz et~al\mbox{.}}{2006}]%
        {Hertz:2006}
\bibfield{author}{\bibinfo{person}{Matthew Hertz}, \bibinfo{person}{Stephen~M.
  Blackburn}, \bibinfo{person}{J.~Eliot~B. Moss}, \bibinfo{person}{Kathryn~S.
  McKinley}, {and} \bibinfo{person}{Darko Stefanovi\'{c}}.}
  \bibinfo{year}{2006}\natexlab{}.
\newblock \showarticletitle{Generating Object Lifetime Traces with Merlin}.
\newblock \bibinfo{journal}{{\em ACM Trans. Program. Lang. Syst.\/}}
  \bibinfo{volume}{28}, \bibinfo{number}{3} (\bibinfo{date}{May}
  \bibinfo{year}{2006}), \bibinfo{pages}{476--516}.
\newblock
\showISSN{0164-0925}
\showURL{%
\url{http://doi.acm.org/10.1145/1133651.1133654}}


\bibitem[\protect\citeauthoryear{Jones, Hosking, and Moss}{Jones
  et~al\mbox{.}}{2016}]%
        {Jones:2016}
\bibfield{author}{\bibinfo{person}{Richard Jones}, \bibinfo{person}{Antony
  Hosking}, {and} \bibinfo{person}{Eliot Moss}.}
  \bibinfo{year}{2016}\natexlab{}.
\newblock \bibinfo{booktitle}{{\em The garbage collection handbook: the art of
  automatic memory management}}.
\newblock \bibinfo{publisher}{CRC Press}.
\newblock


\bibitem[\protect\citeauthoryear{Jones and Ryder}{Jones and Ryder}{2008}]%
        {Jones:2008}
\bibfield{author}{\bibinfo{person}{Richard~E. Jones} {and}
  \bibinfo{person}{Chris Ryder}.} \bibinfo{year}{2008}\natexlab{}.
\newblock \showarticletitle{A Study of Java Object Demographics}. In
  \bibinfo{booktitle}{{\em Proceedings of the 7th International Symposium on
  Memory Management}} {\em (\bibinfo{series}{ISMM '08})}.
  \bibinfo{publisher}{ACM}, \bibinfo{address}{New York, NY, USA},
  \bibinfo{pages}{121--130}.
\newblock
\showISBNx{978-1-60558-134-7}
\showDOI{%
\url{https://doi.org/10.1145/1375634.1375652}}


\bibitem[\protect\citeauthoryear{Kwak, Lee, Park, and Moon}{Kwak
  et~al\mbox{.}}{2010}]%
        {Kwak:2010}
\bibfield{author}{\bibinfo{person}{Haewoon Kwak}, \bibinfo{person}{Changhyun
  Lee}, \bibinfo{person}{Hosung Park}, {and} \bibinfo{person}{Sue Moon}.}
  \bibinfo{year}{2010}\natexlab{}.
\newblock \showarticletitle{What is Twitter, a Social Network or a News
  Media?}. In \bibinfo{booktitle}{{\em Proceedings of the 19th International
  Conference on World Wide Web}} {\em (\bibinfo{series}{WWW '10})}.
  \bibinfo{publisher}{ACM}, \bibinfo{address}{New York, NY, USA},
  \bibinfo{pages}{591--600}.
\newblock
\showISBNx{978-1-60558-799-8}
\showDOI{%
\url{https://doi.org/10.1145/1772690.1772751}}


\bibitem[\protect\citeauthoryear{Kyrola, Blelloch, and Guestrin}{Kyrola
  et~al\mbox{.}}{2012}]%
        {Kyrola:2012}
\bibfield{author}{\bibinfo{person}{Aapo Kyrola}, \bibinfo{person}{Guy
  Blelloch}, {and} \bibinfo{person}{Carlos Guestrin}.}
  \bibinfo{year}{2012}\natexlab{}.
\newblock \showarticletitle{GraphChi: Large-scale Graph Computation on Just a
  PC}. In \bibinfo{booktitle}{{\em Proceedings of the 10th USENIX Conference on
  Operating Systems Design and Implementation}} {\em
  (\bibinfo{series}{OSDI'12})}. \bibinfo{publisher}{USENIX Association},
  \bibinfo{address}{Berkeley, CA, USA}, \bibinfo{pages}{31--46}.
\newblock
\showISBNx{978-1-931971-96-6}
\showURL{%
\url{http://dl.acm.org/citation.cfm?id=2387880.2387884}}


\bibitem[\protect\citeauthoryear{Lakshman and Malik}{Lakshman and
  Malik}{2010}]%
        {Lakshman:2010}
\bibfield{author}{\bibinfo{person}{Avinash Lakshman} {and}
  \bibinfo{person}{Prashant Malik}.} \bibinfo{year}{2010}\natexlab{}.
\newblock \showarticletitle{Cassandra: A Decentralized Structured Storage
  System}.
\newblock \bibinfo{journal}{{\em SIGOPS Oper. Syst. Rev.\/}}
  \bibinfo{volume}{44}, \bibinfo{number}{2} (\bibinfo{date}{April}
  \bibinfo{year}{2010}), \bibinfo{pages}{35--40}.
\newblock
\showISSN{0163-5980}
\showDOI{%
\url{https://doi.org/10.1145/1773912.1773922}}


\bibitem[\protect\citeauthoryear{Maas, Asanovi\'{c}, Harris, and
  Kubiatowicz}{Maas et~al\mbox{.}}{2016}]%
        {Maas:2016}
\bibfield{author}{\bibinfo{person}{Martin Maas}, \bibinfo{person}{Krste
  Asanovi\'{c}}, \bibinfo{person}{Tim Harris}, {and} \bibinfo{person}{John
  Kubiatowicz}.} \bibinfo{year}{2016}\natexlab{}.
\newblock \showarticletitle{Taurus: A Holistic Language Runtime System for
  Coordinating Distributed Managed-Language Applications}. In
  \bibinfo{booktitle}{{\em Proceedings of the Twenty-First International
  Conference on Architectural Support for Programming Languages and Operating
  Systems}} {\em (\bibinfo{series}{ASPLOS '16})}. \bibinfo{publisher}{ACM},
  \bibinfo{address}{New York, NY, USA}, \bibinfo{pages}{457--471}.
\newblock
\showURL{%
\url{http://doi.acm.org/10.1145/2872362.2872386}}


\bibitem[\protect\citeauthoryear{Mao, Zhang, and Shen}{Mao
  et~al\mbox{.}}{2009}]%
        {Mao:2009}
\bibfield{author}{\bibinfo{person}{Feng Mao}, \bibinfo{person}{Eddy~Z. Zhang},
  {and} \bibinfo{person}{Xipeng Shen}.} \bibinfo{year}{2009}\natexlab{}.
\newblock \showarticletitle{Influence of Program Inputs on the Selection of
  Garbage Collectors}. In \bibinfo{booktitle}{{\em Proceedings of the 2009 ACM
  SIGPLAN/SIGOPS International Conference on Virtual Execution Environments}}
  {\em (\bibinfo{series}{VEE '09})}. \bibinfo{publisher}{ACM},
  \bibinfo{pages}{91--100}.
\newblock
\showISBNx{978-1-60558-375-4}
\showURL{%
\url{http://doi.acm.org/10.1145/1508293.1508307}}


\bibitem[\protect\citeauthoryear{McCandless, Hatcher, and
  Gospodnetic}{McCandless et~al\mbox{.}}{2010}]%
        {Mccandless:2010}
\bibfield{author}{\bibinfo{person}{Michael McCandless}, \bibinfo{person}{Erik
  Hatcher}, {and} \bibinfo{person}{Otis Gospodnetic}.}
  \bibinfo{year}{2010}\natexlab{}.
\newblock \bibinfo{booktitle}{{\em Lucene in Action, Second Edition: Covers
  Apache Lucene 3.0}}.
\newblock \bibinfo{publisher}{Manning Publications Co.},
  \bibinfo{address}{Greenwich, CT, USA}.
\newblock
\showISBNx{1933988177, 9781933988177}


\bibitem[\protect\citeauthoryear{Nguyen, Fang, Xu, Demsky, Lu, Alamian, and
  Mutlu}{Nguyen et~al\mbox{.}}{2016}]%
        {Nguyen:2016}
\bibfield{author}{\bibinfo{person}{Khanh Nguyen}, \bibinfo{person}{Lu Fang},
  \bibinfo{person}{Guoqing Xu}, \bibinfo{person}{Brian Demsky},
  \bibinfo{person}{Shan Lu}, \bibinfo{person}{Sanazsadat Alamian}, {and}
  \bibinfo{person}{Onur Mutlu}.} \bibinfo{year}{2016}\natexlab{}.
\newblock \showarticletitle{Yak: A High-performance Big-data-friendly Garbage
  Collector}. In \bibinfo{booktitle}{{\em Proceedings of the 12th USENIX
  Conference on Operating Systems Design and Implementation}} {\em
  (\bibinfo{series}{OSDI'16})}. \bibinfo{publisher}{USENIX Association},
  \bibinfo{address}{Berkeley, CA, USA}, \bibinfo{pages}{349--365}.
\newblock
\showISBNx{978-1-931971-33-1}
\showURL{%
\url{http://dl.acm.org/citation.cfm?id=3026877.3026905}}


\bibitem[\protect\citeauthoryear{Nguyen, Wang, Bu, Fang, Hu, and Xu}{Nguyen
  et~al\mbox{.}}{2015}]%
        {Nguyen:2015}
\bibfield{author}{\bibinfo{person}{Khanh Nguyen}, \bibinfo{person}{Kai Wang},
  \bibinfo{person}{Yingyi Bu}, \bibinfo{person}{Lu Fang},
  \bibinfo{person}{Jianfei Hu}, {and} \bibinfo{person}{Guoqing Xu}.}
  \bibinfo{year}{2015}\natexlab{}.
\newblock \showarticletitle{FACADE: A Compiler and Runtime for (Almost)
  Object-Bounded Big Data Applications}. In \bibinfo{booktitle}{{\em
  Proceedings of the Twentieth International Conference on Architectural
  Support for Programming Languages and Operating Systems}} {\em
  (\bibinfo{series}{ASPLOS '15})}. \bibinfo{publisher}{ACM},
  \bibinfo{pages}{675--690}.
\newblock
\showISBNx{978-1-4503-2835-7}
\showURL{%
\url{http://doi.acm.org/10.1145/2694344.2694345}}


\bibitem[\protect\citeauthoryear{Paleczny, Vick, and Click}{Paleczny
  et~al\mbox{.}}{2001}]%
        {Paleczny:2001}
\bibfield{author}{\bibinfo{person}{Michael Paleczny},
  \bibinfo{person}{Christopher Vick}, {and} \bibinfo{person}{Cliff Click}.}
  \bibinfo{year}{2001}\natexlab{}.
\newblock \showarticletitle{The Java hotspotTM Server Compiler}. In
  \bibinfo{booktitle}{{\em Proceedings of the 2001 Symposium on JavaTM Virtual
  Machine Research and Technology Symposium - Volume 1}} {\em
  (\bibinfo{series}{JVM'01})}. \bibinfo{publisher}{USENIX Association},
  \bibinfo{address}{Berkeley, CA, USA}, \bibinfo{pages}{1--1}.
\newblock
\showURL{%
\url{http://dl.acm.org/citation.cfm?id=1267847.1267848}}


\bibitem[\protect\citeauthoryear{Ricci, Guyer, and Moss}{Ricci
  et~al\mbox{.}}{2011}]%
        {Ricci:2011}
\bibfield{author}{\bibinfo{person}{Nathan~P. Ricci}, \bibinfo{person}{Samuel~Z.
  Guyer}, {and} \bibinfo{person}{J.~Eliot~B. Moss}.}
  \bibinfo{year}{2011}\natexlab{}.
\newblock \showarticletitle{Elephant Tracks: Generating Program Traces with
  Object Death Records}. In \bibinfo{booktitle}{{\em Proceedings of the 9th
  International Conference on Principles and Practice of Programming in Java}}
  {\em (\bibinfo{series}{PPPJ '11})}. \bibinfo{pages}{139--142}.
\newblock
\showISBNx{978-1-4503-0935-6}
\showURL{%
\url{http://doi.acm.org/10.1145/2093157.2093178}}


\bibitem[\protect\citeauthoryear{Sewe, Yuan, Sinschek, and Mezini}{Sewe
  et~al\mbox{.}}{2010}]%
        {Sewe:2010}
\bibfield{author}{\bibinfo{person}{Andreas Sewe}, \bibinfo{person}{Dingwen
  Yuan}, \bibinfo{person}{Jan Sinschek}, {and} \bibinfo{person}{Mira Mezini}.}
  \bibinfo{year}{2010}\natexlab{}.
\newblock \showarticletitle{Headroom-based Pretenuring: Dynamically Pretenuring
  Objects That Live "Long Enough"}. In \bibinfo{booktitle}{{\em Proceedings of
  the 8th International Conference on the Principles and Practice of
  Programming in Java}} {\em (\bibinfo{series}{PPPJ '10})}.
  \bibinfo{publisher}{ACM}, \bibinfo{pages}{29--38}.
\newblock
\showISBNx{978-1-4503-0269-2}
\showURL{%
\url{http://doi.acm.org/10.1145/1852761.1852767}}


\bibitem[\protect\citeauthoryear{Smith}{Smith}{2000}]%
        {Smith:2000}
\bibfield{author}{\bibinfo{person}{Michael~D. Smith}.}
  \bibinfo{year}{2000}\natexlab{}.
\newblock \showarticletitle{Overcoming the Challenges to Feedback-directed
  Optimization (Keynote Talk)}. In \bibinfo{booktitle}{{\em Proceedings of the
  ACM SIGPLAN Workshop on Dynamic and Adaptive Compilation and Optimization}}
  {\em (\bibinfo{series}{DYNAMO '00})}. \bibinfo{publisher}{ACM},
  \bibinfo{pages}{1--11}.
\newblock
\showURL{%
\url{http://doi.acm.org/10.1145/351397.351408}}


\bibitem[\protect\citeauthoryear{Tofte and Talpin}{Tofte and Talpin}{1997}]%
        {Tofte:1997}
\bibfield{author}{\bibinfo{person}{Mads Tofte} {and}
  \bibinfo{person}{Jean-Pierre Talpin}.} \bibinfo{year}{1997}\natexlab{}.
\newblock \showarticletitle{Region-Based Memory Management}.
\newblock \bibinfo{journal}{{\em Inf. Comput.\/}} \bibinfo{volume}{132},
  \bibinfo{number}{2} (\bibinfo{date}{Feb.} \bibinfo{year}{1997}),
  \bibinfo{pages}{109--176}.
\newblock
\showISSN{0890-5401}
\showDOI{%
\url{https://doi.org/10.1006/inco.1996.2613}}


\bibitem[\protect\citeauthoryear{Ungar}{Ungar}{1984}]%
        {Ungar:1984}
\bibfield{author}{\bibinfo{person}{David Ungar}.}
  \bibinfo{year}{1984}\natexlab{}.
\newblock \showarticletitle{Generation Scavenging: A Non-disruptive High
  Performance Storage Reclamation Algorithm}. In \bibinfo{booktitle}{{\em
  Proceedings of the First ACM SIGSOFT/SIGPLAN Software Engineering Symposium
  on Practical Software Development Environments}} {\em (\bibinfo{series}{SDE
  1})}. \bibinfo{publisher}{ACM}, \bibinfo{address}{New York, NY, USA},
  \bibinfo{pages}{157--167}.
\newblock
\showISBNx{0-89791-131-8}
\showDOI{%
\url{https://doi.org/10.1145/800020.808261}}


\bibitem[\protect\citeauthoryear{Vall{\'e}e-Rai, Gagnon, Hendren, Lam,
  Pominville, and Sundaresan}{Vall{\'e}e-Rai et~al\mbox{.}}{2000}]%
        {Vallee-Rai:2000}
\bibfield{author}{\bibinfo{person}{Raja Vall{\'e}e-Rai},
  \bibinfo{person}{Etienne Gagnon}, \bibinfo{person}{Laurie~J. Hendren},
  \bibinfo{person}{Patrick Lam}, \bibinfo{person}{Patrice Pominville}, {and}
  \bibinfo{person}{Vijay Sundaresan}.} \bibinfo{year}{2000}\natexlab{}.
\newblock \showarticletitle{Optimizing Java Bytecode Using the Soot Framework:
  Is It Feasible?}. In \bibinfo{booktitle}{{\em Proceedings of the 9th
  International Conference on Compiler Construction}} {\em (\bibinfo{series}{CC
  '00})}. \bibinfo{publisher}{Springer-Verlag}, \bibinfo{address}{London, UK,
  UK}, \bibinfo{pages}{18--34}.
\newblock
\showISBNx{3-540-67263-X}
\showURL{%
\url{http://dl.acm.org/citation.cfm?id=647476.727758}}


\bibitem[\protect\citeauthoryear{Xu}{Xu}{2013}]%
        {Xu:2013}
\bibfield{author}{\bibinfo{person}{Guoqing Xu}.}
  \bibinfo{year}{2013}\natexlab{}.
\newblock \showarticletitle{Resurrector: A Tunable Object Lifetime Profiling
  Technique for Optimizing Real-world Programs}. In \bibinfo{booktitle}{{\em
  Proceedings of the 2013 ACM SIGPLAN International Conference on Object
  Oriented Programming Systems Languages \&\#38; Applications}} {\em
  (\bibinfo{series}{OOPSLA '13})}. \bibinfo{publisher}{ACM},
  \bibinfo{pages}{111--130}.
\newblock
\showISBNx{978-1-4503-2374-1}
\showURL{%
\url{http://doi.acm.org/10.1145/2509136.2509512}}


\bibitem[\protect\citeauthoryear{Zheng, Bulej, and Binder}{Zheng
  et~al\mbox{.}}{2015}]%
        {Zheng:2015}
\bibfield{author}{\bibinfo{person}{Yudi Zheng}, \bibinfo{person}{Lubom\'{\i}r
  Bulej}, {and} \bibinfo{person}{Walter Binder}.}
  \bibinfo{year}{2015}\natexlab{}.
\newblock \showarticletitle{Accurate Profiling in the Presence of Dynamic
  Compilation}. In \bibinfo{booktitle}{{\em Proceedings of the 2015 ACM SIGPLAN
  International Conference on Object-Oriented Programming, Systems, Languages,
  and Applications}} {\em (\bibinfo{series}{OOPSLA 2015})}.
  \bibinfo{publisher}{ACM}, \bibinfo{address}{New York, NY, USA},
  \bibinfo{pages}{433--450}.
\newblock
\showISBNx{978-1-4503-3689-5}
\showURL{%
\url{http://doi.acm.org/10.1145/2814270.2814281}}


\end{thebibliography}

\end{document}